\documentclass[a4paper,structabstract]{aa}
\usepackage{amsmath}
\usepackage{amssymb}
\usepackage{graphicx}
\usepackage{subfigure}
\usepackage[innercaption]{sidecap}
\usepackage{epsfig}
\usepackage{color}
\usepackage{txfonts}
\usepackage{natbib}
\usepackage[bookmarks=false,colorlinks=true, citecolor=blue, backref=false,breaklinks=true]{hyperref} 
\usepackage{hyperref}
\bibpunct{(}{)}{;}{a}{}{,} 

\newcommand{\GAS}{\texttt{G.A.S.}}
\newcommand{\ds}{\displaystyle}

\newcommand{\Msun}{M_{\odot}}

\newcommand{\Zsun}{Z_{\odot}}

\begin{document}

\title{\GAS\ II: Dust extinction in galaxies;\\ Luminosity functions and InfraRed eXcess}
\titlerunning{\GAS\ II: Dust extinction in galaxies}

\author{M. Cousin \inst{1} \and V. Buat \inst{1} \and G. Lagache\inst{1}  \and M. Bethermin\inst{1}}

\institute{Aix Marseille Univ, CNRS, CNES, LAM, Marseille, France.\\ \email{morgane.cousin86@gmail.com}, website: {\tt http://morganecousin.wordpress.com}} 

\date{Received November 19 2018 / Accepted January 17 2019}

\abstract{Dust is a crucial component of the interstellar medium of galaxies. The presence of dust strongly affects the light produced by stars within a galaxy. As these photons are our main information vector to explore the stellar mass assembly and therefore understand a galaxy's evolution, modeling the luminous properties of galaxies and taking into account the impact of the dust is a fundamental challenge for semi-analytical models.}
{We present the complete prescription of dust attenuation implemented in the new semi-analytical model (SAM): \GAS\ . This model is based on a two-phase medium originating from a physically motivated turbulent model of gas structuring (\GAS\ I paper).}
{Dust impact is treated by taking into account three dust components: Polycyclic Aromatic Hydrocarbons, Very Small Grains, and Big Grains. All three components evolve in both a diffuse and a fragmented/dense gas phase. Each phase has its own stars, dust content and geometry. Dust content evolves according to the metallicity of it associated phase.}
{The \GAS\, model is used to predict both the UV and the IR luminosity functions from z=9.0 to z=0.1. Our two-phase ISM prescription catches very well the evolution of UV and IR luminosity functions. We note a small overproduction of the IR luminosity at low redshift ($z<0.5$). We also focus on the Infrared-Excess (IRX) and explore its dependency with the stellar mass, UV slope, stellar age, metallicity and slope of the attenuation curves. Our model predicts large scatters for relations based on IRX, especially for the IRX-$\beta$ relation. Our analysis reveals that the slope of the attenuation curve is more driven by absolute attenuation in the FUV band than by disk inclination. We confirm that the age of the stellar population and the slope of the attenuation curve can both shift galaxies below the fiducial star-birth relation in the IRX-$\beta$ diagram. Main results presented in this paper (e.g. luminosity functions) and in the two other associated \GAS\, papers are stored and available in the GALAKSIENN library through the ZENODO platform.}
{}

\keywords{Galaxies: formation - Galaxies: evolution - Infrared: galaxies - Ultraviolet: galaxies; Dust: galaxies}

\maketitle

%
%

\section{Introduction}

In galaxies, stars, gas, and dust are mixed and strongly interact. Dust plays a major role in the chemistry and the physics of the interstellar medium (ISM). Whereas stars produce Ultra-Violet (UV) and visible and near-infrared (NIR) light, dust, heated by stellar radiation, radiates from the near-InfraRed (nIR) to the sub-millimeter (submm). Attenuation of the stellar light due to dust dramatically modifies fluxes, colors and spectral energy distributions of galaxies. Measurements of  star formation rates (SFRs) and even stellar mass \citep[e.g.][]{Lo_Faro_2017} are therefore strongly affected. A large fraction of the UV radiation coming from newly formed O and B stars is obscured by dust \citep[e.g.][]{Casey_2014, Burgarella_2013, Takeuchi_2005} and re-emitted mainly in the Mid- and Far-InfraRed (IR) and the sub-millimeter (submm) domain. This process is particularly efficient in dusty star forming galaxies (DSFGs), that can reach IR luminosities up to $10^{13}L_{\odot}$ and thus SFRs of thousands of solar masses per year \citep[e.g.][]{Casey_2012, Vieira_2013, Da_Cunha_2015}.

To take into account the overall star formation activity, both UV and IR contributions must be considered. By combining both  measurements, \cite{Burgarella_2013} inferred the redshift evolution (0 $<$ z $<$ 4) of the total cosmic star formation activity. Overall attenuation affecting the galaxies is often parametrized by the IR excess: $IRX = L_{IR}/L_{UV}$, where $L_{IR}$ is the integrated infrared luminosity between 8 and 1000\,microns and $L_{UV}$ is the rest-frame 1500\,\AA\, luminosity.

A large number of studies have measured the IRX. In the nearby Universe \citep[e.g.][]{Martin_2005b,Bothwell_2011}, half of the total UV budget is obscured. At higher redshift, the fraction of the star formation activity directly traced by the UV strongly decreases, being less than 15\% at $z>1$ \citep[e.g.,][]{Takeuchi_2005,Cucciati_2012}. Due to the decrease of the metallicity in very young galaxies, at $z>$3, it is expected that the UV flux traces a higher fraction of the star formation activity ($<$~20\%, \citealt{Reddy_2008}, \citealt{Bouwens_2010}). Thanks to recent observational progresses, building large UV-selected samples is feasible over most of the evolution of the Universe \citep[e.g.][]{Bouwens_2012, Ellis_2013, Reddy_2012} and the majority of studies are based on (rest-frame) UV selected samples.

At high redshift (z $>$ 3) or in fields where IR/submm measurements are not available, only the stellar emission (UV-NIR) attenuated by dust is observed. Estimating accurately the corrections for this attenuation is thus crucial to derive SFRs. One of the most popular method to derive dust obscuration is based on the observed correlation between the slope of the UV continuum ($\beta$) and IRX. Initially measured and calibrated in nearby starbursts \citep[e.g.][]{Meurer_1999}, this method has been extensively used at all redshifts. However, it was found to be invalid for nearby secular star forming galaxies \cite[e.g.][]{Boissier_2007, Hao_2011}. The relation between $\beta$ and IRX is also affected by selection effects \cite[e.g.][]{Buat_2005} and could also be sensitive to dust properties \cite[e.g.][]{Calzetti_2001, Inoue_2006}, star formation histories \cite[e.g.][]{Kong_2004, Boquien_2012, Forrest_2016} and shape/slope of the attenuation law \citep[e.g.][]{Salmon_2016, Reddy_2016, Bouwens_2016}.

In parallel to the IRX$-\beta$ relation another correlation has been measured between IRX and the stellar mass \citep[e.g.][]{Pannella_2009,Buat_2012,Finkelstein_2012,Heinis_2014,Whitaker_2014}. This correlation is currently not fully understood. In this context, stellar mass could be a crude tracer of the metal enrichment process through the complex star formation histories. Indeed, work by \cite{Cortese_2006}, \cite{Boquien_2009}, or more recently \cite{Pannella_2015} has revealed a correlation between the dust attenuation and the gas-phase metallicity, which is itself correlated with the stellar mass \citep[e.g.][]{Mannucci_2010, Lilly_2013, Zahid_2014}.

Understanding dust attenuation in galaxies is still an important challenge and therefore a fundamental aspect of SAMs. The impact of dust has been studied in SAMs with different methods \citep[e.g.][]{Devriendt_1999, Granato_2000, Baugh_2005, Bower_2006, De_Lucia_2007, Fontanot_2009, Guo_2011, Lacey_2016} and yields to predicted number counts in UV and/or luminosity functions that differ by a factor $\simeq$2. The recent work of \cite{Narayanan_2018}, based on hydrodynamic zoom-in simulations and radiative transfer, and the analytical prescription proposed by \cite{Popping_2017} highlight the complex effects of the average stellar population age, attenuation curves, and dust/stars mixing. However, the analytical model by \cite{Popping_2017} cannot take into account complex star formation histories and, even if the hydrodynamic simulations from \cite{Narayanan_2018} have been performed from z $\simeq$ 5 to z $\simeq$ 2, their limited number of simulated galaxies does not allow to explore the redshift evolution of the IRX-$\beta$ relation.

In this paper (paper II of the \GAS\ model presentation set), we present the set of prescriptions implemented in the \GAS\ model to describe the effects of the dust attenuation onto the stellar light. Physical prescriptions associated to the gas physics is describe in \cite{Cousin_2018a} (paper \GAS\ I).

We investigate in this paper the redshift evolution of the UltraViolet (UV) and InfraRed luminosity functions. We also explore the main drivers of both the IRX-$M_{\star}$ and $IRX$-$\beta$ relations in a realistic cosmological context from z $>$ 6 to z $\simeq$ 1.5. We explore the impact of the stellar age, the star formation rate, the absolute attenuation, the disk inclination and the the gas-phase metallicity. We will question the validity of the IRX-$\beta$ relation to correct UV observed flux from the dust attenuation. 

The paper is organized as follows: in Sect. \ref{sec:model}, we briefly describe our \GAS\ model. In Sect. \ref{sec:dust} we describe dust components. We also present the recipes implemented to take into account dust attenuation and explore the impact of geometry and absolute attenuation. We present the evolution of the attenuation law predicted by our model as function of both the absolute attenuation in FUV band and disk inclination. We also compare our attenuation curves with standard prescriptions. In Sect. \ref{sec:Luminosity_functions} we compare our predictions for the UV and IR luminosity functions to a large set of observational measurements. In Sect. \ref{sec:dust_attenuation_IRX} we focus on correlations between IRX, stellar mass, stellar population age, UV luminosity, gas phase metallicity, UV slope. We conclude in Sect.\,\ref{sec:discussion_conclusion}.  

%
%

\section{The \GAS\ model}
\label{sec:model}

Our work is the continuation of the \GAS\ semi-analytical model described in the \GAS\ I paper \citep{Cousin_2018a}. \GAS\ is applied to dark-matter merger trees extracted from a pure N-body simulation. This simulation uses a WMAP-5yr cosmology ($\Omega_m = 0.28$, $\Omega_{\Lambda} = 0.72$, $f_b = 0.16$, $h = 0.70$ \cite{Komatsu_2009}) and describes a volume of $[100/h]^3 Mpc$ containing $1024^3$ particles. Each of these particles has a mass of $m_p = 1.025~10^8~\Msun$. Dark-matter halos and sub-structures (satellites) are identified by using the \verb?HaloMaker? code \citep{Tweed_2009}. In merger tree structures, we only take into account halos with at least twenty dark-matter particles leading to a minimal dark matter halo mass of $2.050\times10^9~\Msun$.

\subsection{General description}
\label{sec:model_short_description}

Galaxy assembly starts from gas accretion. In \GAS halos are fed through a cosmological smooth metal-free baryonic accretion following the smooth dark-mater accretion. The total accreted gas is divided in two channels -- a cold and a hot mode -- depending on the dark-matter halo mass. While in the cold mode the gas directly feeds the galaxies, the shock-heated gas of the hot mode has to cool down and condensate before feeding the galaxy (\GAS\ I paper, Sect. 2). We assume that the warm condensed gas feeds a gaseous disk with an exponential profile. Gas acquires angular momentum during the transfer \citep{Peebles_1969} and the disk is then supported by this angular momentum \citep[e.g.][]{Blumenthal_1986, Mo_1998}.

In the disk the accreted gas is initially assumed to be mainly diffuse and non-star-forming. It goes through a progressive fragmentation following a turbulent cascade and is finally converted in star-forming gas. Galaxies modelled by \GAS\ host therefore two distinct ISM phases: a diffuse warm phase and a fragmented dense cold phase (\GAS\ I paper, Sect. 3 and 4). Young stars are born in the dense, structured gas. We assume that they stay in this dense phase during $5\times 10^{7}$ yrs. Older stars then evolve in the diffuse warm phase of the disk. 

Our model takes into account supernovae (SNs) and active galaxy nuclei (AGN) feedback. A part of the power produced by these two types of feedback is directly injected in the interstellar medium of the galaxy and is used to disrupt the fragmented dense gas. The  remaining power allows to generate large-scale ejecta and heat the gas. The gas ejected from the galaxy ends up in the hot gas atmosphere surrounding the galaxy. These feedback mechanisms mainly affect the evolution of low-mass and intermediate-mass galaxies ($<10^{9.5}\Msun$) (\GAS\ I paper, Sect. 5). 

In addition to these processes, we introduce a new regulation mechanism operating in the surrounding hot gas phase, that allows to reduce gas cooling in massive halos. It is based on thermal instabilities (\GAS\ I paper, Sect. 6).

Using these new gas-cycle and gas-regulation prescriptions, the \GAS\ model is able to reproduce the stellar mass function of galaxies from z $\simeq$ 6 to z$\simeq$0.5.

\subsection{The chemodynamical model}
\label{sec:chemo_prescription}

A chemodynamical model \citep{Cousin_2016} is used to track the mass evolution (e.g. $M_O(t)$) of six of the main ISM elements: $^1H$, $^4He$, $^{12}C$, $^{14}N$, $^{16}O$, and $^{56}Fe$. Masses are tracked separately: on one side in the diffuse and the fragmented/dense gas phase of the disc, on the other side in the diffuse gas phase of the central bulge (if any).  The production and the re-injection of these elements are taken into account for stars with initial mass between 0.1$\Msun$ and 100$\Msun$ and for metal-free to super-solar metal fractions. We assume that new stars are formed in the dense/fragmented gas phase following a \cite{Chabrier_2003} initial mass function (IMF). Metals are injected initially into the dense/fragmented gas phase. The kinetic energy coming from both SNs and AGNs partly disrupts the dense/fragmented gas phase and injects metals in the diffuse gas phase. Then metals evolving in both the diffuse and the dense/fragmented gas are strewn in the hot atmosphere surrounding the galaxy with the large scale ejecta. Due to this metal enrichment, the efficiency of the radiative cooling process affecting the hot surrounding gas is modified and impact the gas accretion.

Oxygen is the most abundant element formed into stars. It is commonly used as a tracer of the gas-phase metallicity. We define the gas metallicity $Z_g$ of a given phase as the number of oxygen atoms per hydrogen atoms with a logarithmic scale: 
\begin{equation}
	Z_g = 12 + log(O/H) = 12 + log_{10}\left(\dfrac{M_O}{M_H}\dfrac{m_H}{m_O}\right)
	\label{eq:gas_metallicity}
\end{equation}
$M_O$ and $M_H$ are the oxygen and the hydrogen mass contained in a given gas phase. $m_O$ and $m_H$ are the atomic mass of oxygen and hydrogen, respectively. In this formalism we adopt $\Zsun = 8.94$ \citep{Karakas_2010}.

\subsection{Stellar spectra}
\label{sec:stellar_spectra}
Similarly to the gas metallicity, the \GAS\ model tracks the evolution of the spectral energy distribution (SED) of stellar populations. The star formation history of a galaxy component (disk or central bulge) is continuously followed according to the stellar age and the initial stellar metallicity. Thus, at each step of the evolution, we associate to a given stellar population its mass-weighted stellar SED: $(\lambda I_{\lambda})_{\star}$ based on \cite{Bruzual_2003} SED libraries.

\subsection{Merger events}
\label{sec:merger_events}

We define the merger type (major or minor) of two galaxies 1 and 2 using the following mass ratio: 
\begin{equation}
    \eta_{m} = \dfrac{MIN(M_{1/2,1}~;~M_{1/2,2})}{MAX(M_{1/2,1}~;~M_{1/2,2})}\,,
    \label{eq:eta_merger}
\end{equation}
where $M_{1/2}$ is the total mass (galaxy and dark-matter halo) inside the galaxy half-mass radius. We consider the merger event as minor when $\eta_{m} < 1/3$.

In the case of a major merger ($\eta_{m} \ge 1/3$), the galaxy structure and the dynamics are strongly modified. All the gas (from the disk and the bulge) is attributed to the remnant disk. 

We assume that young stars ($\le 5\times 10^7$yrs) of the two progenitor disks are still embedded in GMCs. Young stars are therefore kept into the remnant disk. 

Conversely, old stars ($> 5\times 10^7$yrs) coming from the two progenitors are transferred to the central bulge. The stellar mass distribution of the central bulge is modeled by a \cite{Hernquist_1990} profile. 

Between two merger events, all the gas produced by stellar winds from stars in the bulge is kept in the bulge. The mass that is generated is in average low ($M_g < 10^5 \Msun$) but the metallicity can be relatively high ($Z_g > 8.0$).

During a minor merger, we add separately the stellar population of the bulges and disks to form the remnant bulge/disk. All gas reservoirs are added into the remnant disk.

%
%

\section{Dust in galaxies}
\label{sec:dust}

In a galaxy, the overall extinction is the result of different factors: i) the amount/composition of dust in the different components of the galaxy (disk, bulge); ii) the relative contribution of the various stellar populations (in age and metallicity) in the various components; iii) the relative geometry of the dust and stars.

\subsection{Dust composition}

\begin{table}[h]
  \begin{center}
    \footnotesize{
      \begin{tabular}{l|c|c|c}
        \hline
        Type & $\rho$ [$g/cm^3$] & size range [cm] & size law \\
        \hline     
        \hline
        PAH0     & $2.24$  & $3.1\times 10^{-8}$ - $1.2\times 10^{-7}$ & mix-logn \\
		PAH1     & $2.24$  & $3.1\times 10^{-8}$ - $1.2\times 10^{-7}$ & mix-logn \\
		\hline  		
		\hline		
		amCBE  & $1.81$  & $6.0\times 10^{-8}$ - $2.0\times 10^{-6}$ & logn \\         
	 	amCBE  & $1.81$  & $4.0\times 10^{-7}$ - $2.0\times 10^{-4}$ & plaw-ed \\
        \hline  
		\hline	
		aSil     & $3.00$  & $4.0\times 10^{-7}$ - $2.0\times 10^{-4}$ & plaw-ed \\
        \hline
      \end{tabular}}
  \end{center}  
  \caption{\footnotesize{Dust properties. The dust type and size law refer to {\tt{DustEM}} keywords.}}
  \label{tab:dust_properties}
\end{table} 

As proposed by for example \cite{Draine_Li_2001}, \cite{Zubko_2004},\cite{Draine_Li_2007}, \cite{Compiegne_2011} (and references therein), dust can be described by a set of three different grain types: Polycyclic Aromatic Hydrocarbons (PAH), Very Small Grains (VSG) and Big Grains (BG). In \GAS\, we assume that these three dust types are formed in all galaxies at all redshifts. However, the relative mass fraction of this three dust types can vary from one environment to another (Sect. \ref{sec:dust_contents}). The intrinsic properties of dust assumed in our model are listed in Table~\ref{tab:dust_properties}. These properties are fully compatible to those used in previous work \citep[e.g.][]{Zubko_2004, Draine_Li_2007, Compiegne_2011}.

\subsection{Dust content}
\label{sec:dust_contents}

In the interstellar medium, gas and dust are closely linked and the dust composition (PAH:$f_{PAH}$, Big Grains:$f_{BG}$, Very Small Grains:$f_{VSG}$) depends on the gas metallicity. 

\begin{figure}[t]
	\begin{center}
		\includegraphics[scale=0.7]{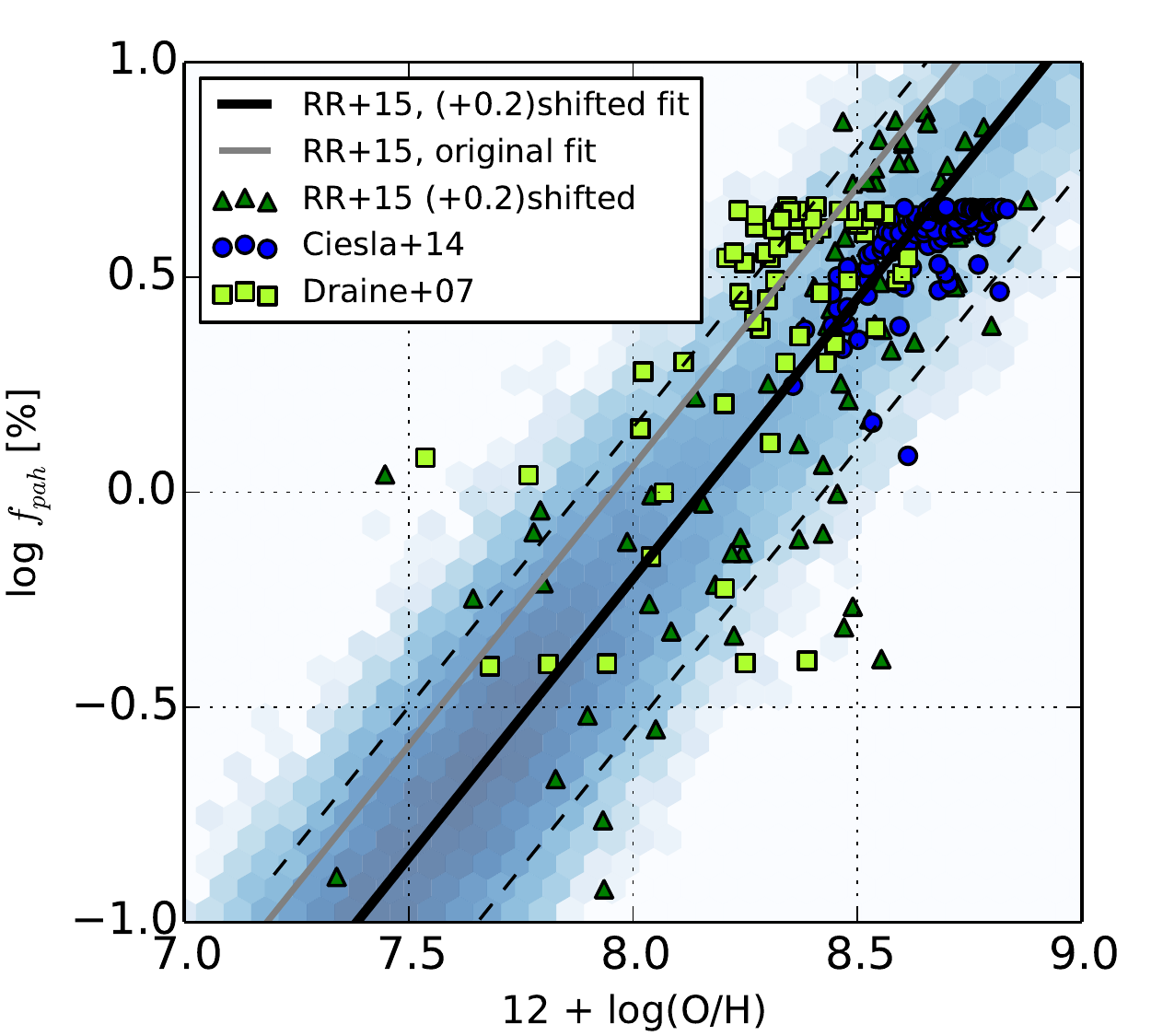}
    	\caption{Fraction of PAH as a function of the gas metallicity in the dense/fragmented gas phase. The blue shaded area shows the full \GAS\ distribution. Blue circles and light-green squares show \cite{Ciesla_2014} and \cite{Draine_2007} data points respectively. Green triangles show \cite{Remy_Ruyer_2015} sample shifted by 0.2\,dex. The black solid line corresponds to our mean relation (Eq. \ref{eq:f_PAH}). The original mean relation of \cite{Remy_Ruyer_2015} (their Eq. 5) is shown with the solid gray line. Dashed black lines mark the $3\sigma$ limits around our mean relation.}
    	\label{fig:f_pah_Zg}
	\end{center}
\end{figure} 

Fig. \ref{fig:f_pah_Zg} shows measurements of the PAH fractions as a function of the gas metallicity performed by \cite{Draine_2007}, \cite{Ciesla_2014} and \cite{Remy_Ruyer_2015}. 

These studies used different prescriptions for gas metallicity measurements. As shown by \cite{Kewley_2008}, the prescription proposed by \citet{Pilyugin_2005} (used in \cite{Remy_Ruyer_2015}) leads to lower values than all other prescriptions. A systematic shift of +0.2 dex applied to \cite{Remy_Ruyer_2015} measurements leads to a good agreement between the studies. 

In the two distinguished ISM phases hosted by our modeled galaxies, the fraction of PAH formed in a given gas-phase is settled by using \cite{Remy_Ruyer_2015} (their Eq. 5): 
\begin{equation}
	\dfrac{f_{PAH}}{f_{PAH,0}} = 10^{-11.0 + 1.30Z_g(-0.2)}
	\label{eq:f_PAH}
\end{equation}
where $f_{PAH,0} = 4.57\%$ is the reference value for our Galaxy \citep{Zubko_2004}. 

Eq. \ref{eq:f_PAH} is used to generate a continuous and smooth evolution of the PAH fraction in relation to the gas phase metallicity. Around the mean relation, we introduce a scatter by applying between two time-steps a random walk between the current value and a target value given by Eq. \ref{eq:f_PAH} (for more details, see Appendix \ref{sec:PAH_random_walk}). We plot in Fig. \ref{fig:f_pah_Zg} the whole distribution of $f_{pah}$ obtained following this random walk procedure and measured in the dense/fragmented gas phase. The scatter obtained around the mean relation is fully compatible with observational measurements.

BG represent the main part of the dust mass \citep[e.g.][]{Zubko_1996, Draine_Li_2001, Compiegne_2011}, therefore when the PAH fraction is settled, we then treat the BG mass fraction. This fraction is assumed to be around to $2/3$ of the residual mass. This fraction is typical of what is measured in our Galaxy \citep[e.g.][]{Zubko_2004, Compiegne_2011}. As for PAH, we computed the fraction of BG following the random-walk technique. At each time step, the mass fraction of BG is randomly settled between the current value and the target value ($2/3$ of the residual mass). Finally, mass conservation requires that VSGs are made of the rest of the dust mass.

The dust composition of each gas phase (diffuse gas in disc and bulge, dense/fragmented gas in disc), is derived independently according to its metallicity, using Eq.\ref{eq:f_PAH} and our random-walk technique (Eq. \ref{eq:gas_metallicity}).

The empirical relation described by Eq. \ref{eq:f_PAH} cannot fully account for the complex dust processing in the ISM: fragmentation, destruction by UV photons, heating, or interstellar shocks \citep{Jones_1994, Jones_1996, Draine_Li_2007, Relano_2018}. When $f_{PAH}$ increases or decreases between two time-steps, according to Eq. \ref{eq:f_PAH} and the random-walk technique, the chosen "effective evolution" hides the complexity, which cannot be treated in full details by \GAS\. Indeed, even if \GAS tracks the kinetic energy produced by SNs and AGN, this energy is only used to (re-)distribute the gas between the different gas phases and does not affects directly the dust contents/composition.

Moreover, \GAS assumes a single relation characterized by a smooth and continuous trend from low to high metallicities. As pointed by \cite{Remy_Ruyer_2015} or \cite{Draine_Li_2007}, the PAH fraction is low in dwarf galaxies and can be larger in more evolved galaxies. The $Z_g$--$f_{PAH}$ relation assumed in \GAS (Eq. \ref{eq:f_PAH}) follow the trend of the majority of current observational measurements. We show in Appendix \ref{sec:PAHs} that the variations of this relation have only a small impact onto the dust attenuation. 

\subsection{Extinction curves}
\label{sec:dust_extinction_curves}

Once the dust composition ($f_{PAH}$, $f_{BG}$, $f_{BG}$) is fixed in the two ISM phases, we build the associated extinction curve. For that, we take into account absorption and scattering properties of dust through the {\tt DustEM} model \citep{Compiegne_2011}.

\begin{figure*}
\begin{center}
\hspace{0.5cm}
\subfigure[Absorption and scattering]{\label{fig:dust_abs_sca}\centering\includegraphics[scale=0.63]{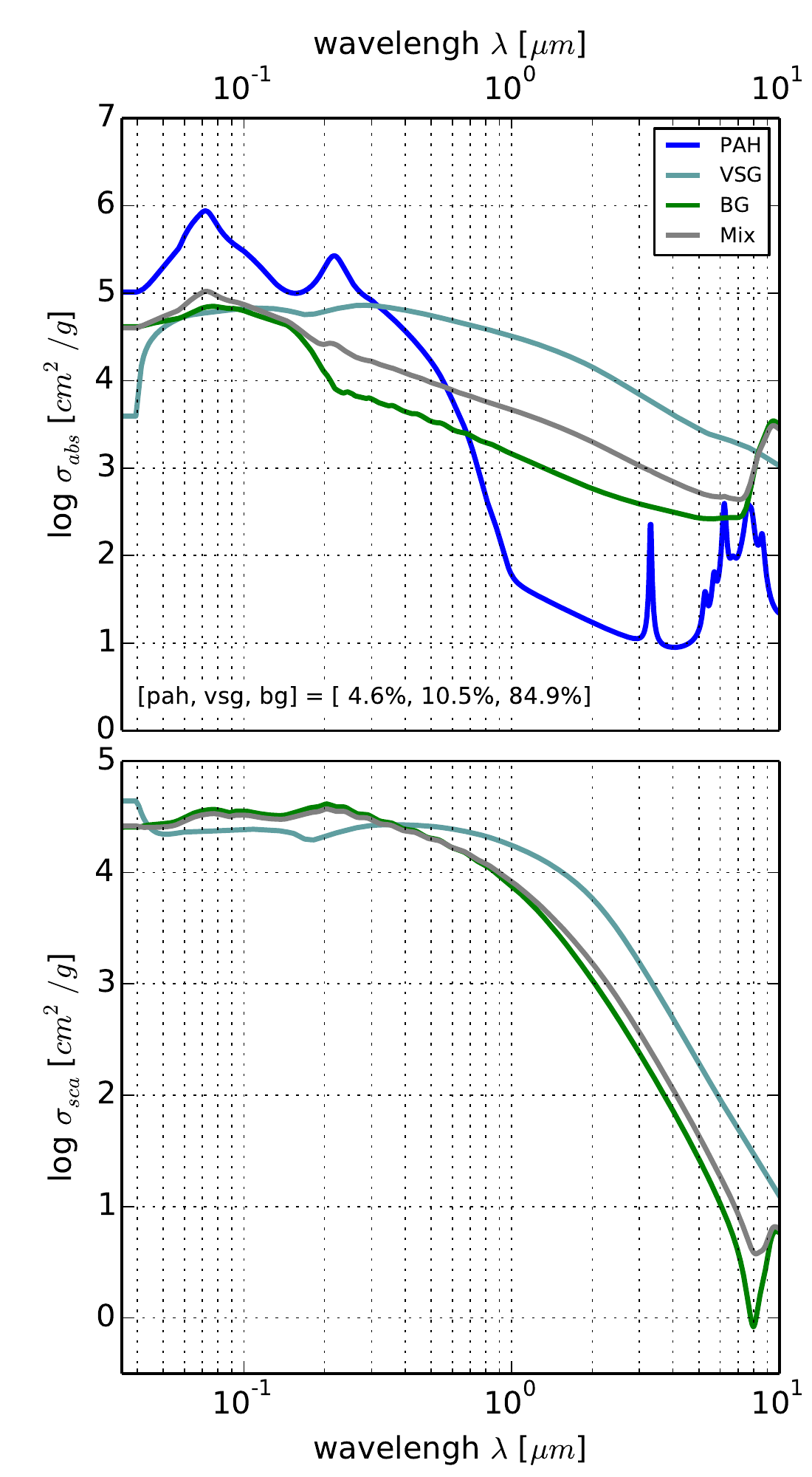}}
\hspace{1.5cm}
\subfigure[Extinction and Albedo]{\label{fig:dust_ext_alb}\includegraphics[scale=0.63]{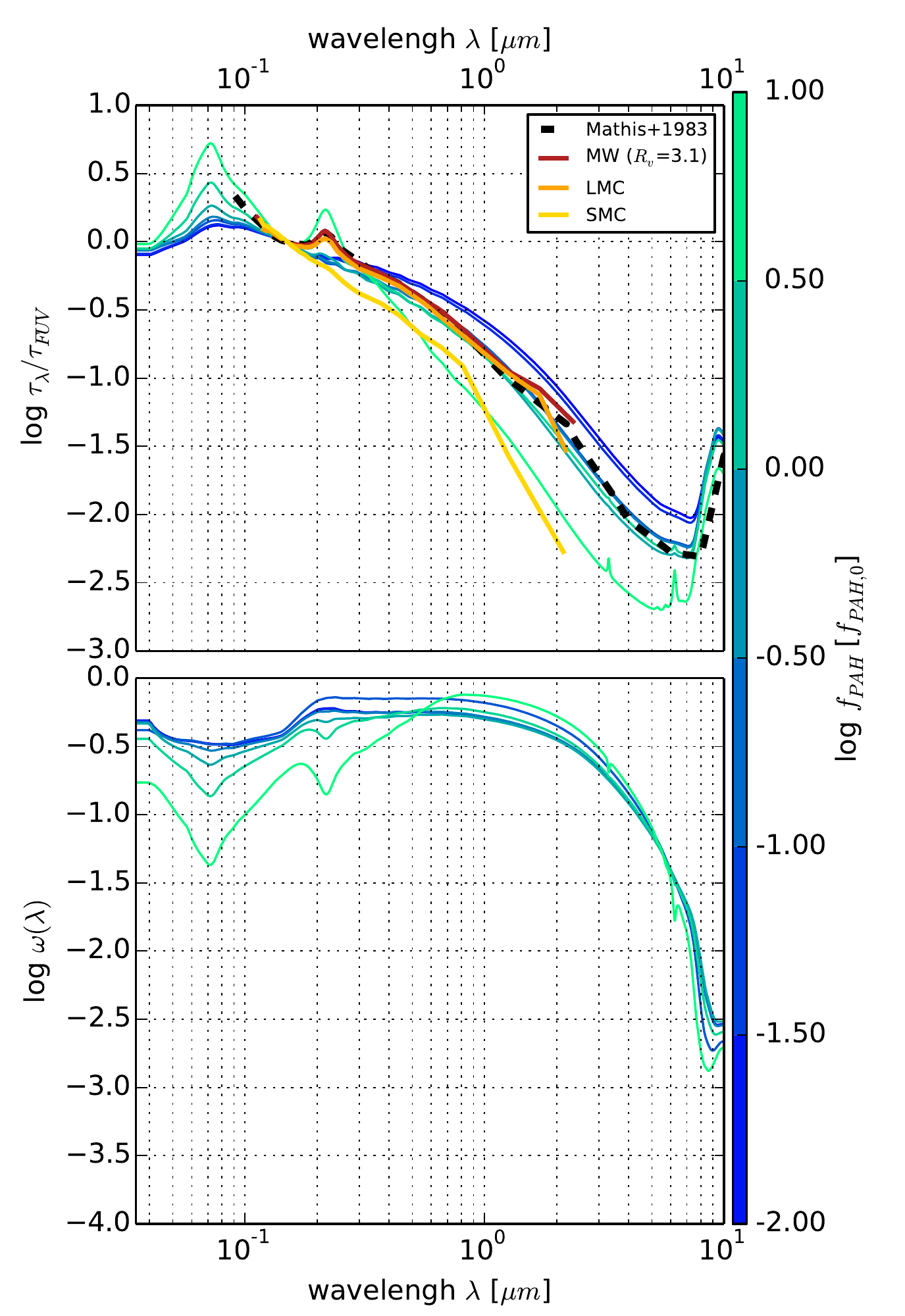}}
\caption{(a): Absorption ($\sigma_{abs}$, upper panel) and scattering ($\sigma_{sca}$ cross sections, lower panel). In each panel, PAH, VSG and BG are shown respectively in blue, light-blue and green, respectively. The gray curve shows the total absorption. We assume $f_{PAH} = f_{PAH,0} = 4.57\%$. The mass fraction of BG and VSG (given in the left panel) are then deduced using the random-walk technique. (b) Extinction ($\tau_{\lambda}$) curves normalized to the FUV band. We compare those extinction curves to observational measurements: \cite{Mathis_1983} Milky-Way (dashed black line) and \cite{Gordon_2003} for SMC (gold solid line), LMC (orange solid line) and Milky-Way (red solid line). Lower Panel: Albedo ($\omega_{\lambda}$). In the two panels, the color of the curves is linked to the PAH fraction assumed, from $10^{-2}\times f_{PAH,0}$ (in blue) to $10\times f_{PAH,0}$ (in green).}
\end{center}
\end{figure*}

Fig. \ref{fig:dust_abs_sca} shows the absorption and the scattering curves associated to the different dust types. 

We define the extinction curve $\tau_\lambda$ as:
\begin{equation}
	\ds\tau_{\lambda} = \sum_{gt} f_{gt}\times \left(\sigma_{abs,gt} + \sigma_{sca,gt} \right)
	\label{eq:extinction_tau_lambda}
\end{equation}
The sum is performed over the three grain types and $f_{gt}$ is the mass fraction associated to a given grain type (see Sect. \ref{sec:dust_contents}).

Based on the scattering and extinction curves, we define the albedo $\omega(\lambda)$,
\begin{equation}
	\ds\omega_{\lambda} = \dfrac{\sum_{gt} f_{gt}\times \sigma_{sca,gt}}{\tau_{\lambda}} \,.
	\label{eq:albedo_omega_lambda}
\end{equation}

Fig. \ref{fig:dust_ext_alb} shows the extinction and albedo wavelength dependencies computed for different PAH fractions. The set of extinction curves produced by \GAS\ is compared to LMC, SMC and Milky Way extinction curves measured by \cite{Mathis_1983} and \cite{Gordon_2003}. In the wavelength range $\lambda\in[0.1, 2.0]\,\mu$m, according to the PAH abundance, our model allows to reproduce a large range of slopes. However, our assumption about the compositions of dust, PAH, VSG and BG leads especially to Milky-Way like extinction curves. 

As highlighted in our Fig. \ref{fig:dust_ext_alb}, both the extinction ($\tau_{\lambda}$) and the albedo ($\omega(\lambda))$ curves are very sensitive to dust composition/properties. These effects are discussed in the literature \citep[e.g.][]{Seon_2016, Chastenet_2017} and are strongly linked to the choices of dust content/composition. They have to be kept in mind.



\subsection{Overall attenuation, dust geometrical distribution}

In a galaxy, the overall attenuation affecting the stellar radiation is governed by the dust content but also by the geometrical distribution of the dust inside the galaxy. On the one hand, the impact of the dust content is quantified by the face-on FUV/optical depth. On the other hand, the geometrical distribution of stars and dust is taken into account through different geometrical functions $\Phi_{\lambda}(\tau_{\lambda},\omega_{\lambda})$. $\Phi_{\lambda}$ allows to build the effective attenuation: 
\begin{equation}
	A_{\lambda} = -2.5~log_{10} \Phi_{\lambda}(\tau_{\lambda},\omega_{\lambda})
\end{equation}
through the extinction curve $\tau_{\lambda}$ and the albedo curve $\omega_{\lambda}$. We define in the next sections different geometrical functions $\Phi_{\lambda}$ which are associated to the different environments/ISM-phases or galaxy components (disk/central bulge) of our modeled galaxies. 

In \cite{Devriendt_1999} or \cite{De_Lucia_2007} the effective attenuation is calibrated according to the Visible (V) band, which is linked to both gas surface density and metallicity (\cite{Guiderdoni_Rocca_1987}). However, the analysis done in \cite{Boquien_2013} (their Fig. 2) indicates that attenuation, measured in the Far Ultra-Violet (FUV) band, is more reliable than in the V-band. They show that the relative scatter of the attenuation in the FUV-band is 30\% to 50\% smaller than in the V-band. These results are only valid for disk galaxies and we thus applied a FUV-band normalization based on \cite{Boquien_2013} to the disks only. We kept the V-band scaling for the bulges. 

\subsubsection{Contents and morphologies of galaxies; impact on the effective attenuation.}
\label{sec:galaxy_environments}

In a given galaxy, the gas/dust distribution is strongly linked to the morphology. Each galaxy is composed of a disk and a potential central bulge. In each disk, the \GAS\ model follows the evolution of two main gas phases: a diffuse warm gas hosting the oldest stars ($> 5\times 10^7$ yrs) and $N_{GMC}$ Giant Molecular Clouds hosting the youngest stars. 

Three different geometrical functions are associated to these three different environments. For the bulge component we adopt a {\tt Dwek} geometry (Sect. \ref{sec:Dwek_geometry}). We use the classical {\tt Slab} (Sect. \ref{sec:Slab_geometry}) geometry for the diffuse ISM. In addition to this {\tt Slab} geometry, young stars evolving in GMCs are described with a {\tt screen} geometry (Sect. \ref{sec:additional_extinction_BC}). In all these cases, we assume that the dust content can affect the stellar radiation of the component only if the gas metallicity is sufficiently high: 12 + log(O/H) $>$ 6.5, else we neglect the dust attenuation.

\subsubsection{Bulge environments: the {\tt Dwek} geometry}
\label{sec:Dwek_geometry}

As explained in Sect. \ref{sec:merger_events}, during a major merger old stars are transferred into a central bulge. Some gas, with metals, is produced by stellar evolution into the bulge. dust formed in this gas phase generate an extinction of the stellar light. We assume that stars and dust are homogeneously mixed and we use the {\tt Dwek} geometrical function \citep{Lucy_1989, Dwek_1996, Devriendt_1999}:
\begin{equation}
	\Phi_{\lambda}^{SPH} = \dfrac{a_{\lambda}}{1 - \omega_{\lambda} + \omega_{\lambda}a_{\lambda}} \,,
	\label{eq:Phi_Dwek_geometry}
\end{equation}
where 
\begin{equation}
	a_{\lambda} = \dfrac{3}{4k_{\lambda}}\left[1-\dfrac{1}{2k^2_{\lambda}}+\left(\dfrac{1}{k_{\lambda}}+\dfrac{1}{2k^2_{\lambda}}\right)exp(-2k_{\lambda})\right] \,,
	\label{eq:a_lambda_Dwek_geometry}
\end{equation}
and 
\begin{equation}
	k_{\lambda} = \tau^{SPH}_{\lambda} = \tau^{SPH}_{V}\times\dfrac{\tau_{\lambda}}{\tau_{\lambda}(\lambda_V)} \,.
	\label{eq:k_lambda_Dwek_geometry}
\end{equation}
$\tau^{SPH}_{V}$, the reference depth in (V-band), varies with the metal mass fraction\footnote{We assume a solar metal mass fraction: $\chi_{\odot} = 0.02$} $\chi_Z$ and the hydrogen column density $N_H$ following \citep{Guiderdoni_Rocca_1987, Devriendt_1999}:
\begin{equation}
	\tau^{SPH}_{V} = 2.619\left(\dfrac{\chi_Z}{\chi_{\odot}}\right)^{1.6}\left(\dfrac{N_H}{2.1\times 10^{21} at\cdot cm^{-2}}\right)
	\label{eq:extinction_tau_lambda_Dwek_geometry}
\end{equation}
The hydrogen column density is computed within the half mass radius and we assume an homogeneous gas metallicity in the bulge.

\subsubsection{disk diffuse ISM environments: the {\tt Slab} geometry}
\label{sec:Slab_geometry}

To describe the dust attenuation in the diffuse warm ISM of each galaxy disk we use the standard {\tt Slab} geometry. This distribution is defined as an infinite plane-parallel in which stars, gas and dust are perfectly mixed over the same scale. We use:
\begin{equation}
	\Phi_{\lambda}^{ISM} = \dfrac{1 - exp\left(-\sqrt{1-\omega_{\lambda}^{ISM}}\tau_{\lambda}^{ISM}/cos~i\right)}{\sqrt{1-\omega_{\lambda}^{ISM}}\tau_{\lambda}^{ISM}/cos~i}
	\label{eq:Phi_Slab_geometry}
\end{equation}
where $i$ is the inclination angle of the disk with respect to the $z$ axis of the simulated box. $\tau_{\lambda}^{ISM}$ is defined as followed:
\begin{equation}
	\tau_{\lambda}^{ISM} = \tau_{FUV}^{ISM}\times\dfrac{\tau_{\lambda}}{\tau_{\lambda}(\lambda_{FUV})}
	\label{eq:extinction_tau_lambda_Slab_geometry}
\end{equation}
The extinction curve $\tau_{\lambda}$ is built according to the dust composition (see Sect. \ref{sec:dust_contents}, Eq. \ref{eq:extinction_tau_lambda}) of the diffuse ISM. This composition depends of the diffuse gas phase metallicity $Z_g^{ISM}$. The face on UV depth $\tau_{FUV}^{ISM}$ is calculated as in \cite{Boquien_2013} (their Eq. 13):
\begin{equation}
	\tau_{FUV}^{ISM} = \left(1.926 + 0.051\times \Sigma_H^{ISM}\right)\times 10^{[0.947\times Z_g^{ISM} -9]}
	\label{eq:tau_UV_Slab}
\end{equation}
To compute the hydrogen surface density $\Sigma_H^{ISM}$ we assume that half of the hydrogen mass, is stored in the half mass radius of the galaxy. $Z_g^{ISM}$ is settled to the gas-phase metallicity specifically associated with the diffuse warm gas contained into the disk. In Eq. \ref{eq:Phi_Slab_geometry}, the albedo is taken into account through the factor $\sqrt{1-\omega_{\lambda}}$.

\subsubsection{Additional attenuation from Giant Molecular Clouds}
\label{sec:additional_extinction_BC}

New stars are embedded in some GMCs. These dense structures generate an additional attenuation that we model by applying a {\tt Screen} geometrical function: 
\begin{equation}
	\Phi_{\lambda}^{GMC} = exp\left(-\sqrt{1-\omega_{\lambda}}\tau_{\lambda}^{GMC}\right)
	\label{eq:Phi_BC}
\end{equation}
Albedo $\omega_{\lambda}$ and extinction curve $\tau_{\lambda}^{GMC}$ are computed following the dust composition of the structured gas phase. $\tau_{\lambda}^{GMC}$ is normalized to the FUV band thought: 
\begin{equation}
	\tau_{\lambda}^{GMC} = \tau_{FUV}^{GMC}\times\dfrac{\tau_{\lambda}}{\tau_{\lambda}(\lambda_{FUV})}
	\label{eq:extinction_tau_lambda_BC}
\end{equation}
The normalization factor $\tau_{FUV}^{GMC}$ is settled with Eq. \ref{eq:tau_UV_Slab}. For that, all GMC formed in a galaxy are assumed to be similar. For a GMC, the average hydrogen surface density is computed by assuming that all the hydrogen mass is stored in $N_{GMC}$ spherical clouds of radius $R_{GMC}=3h/8$ (see \cite{Cousin_2018a}). Normalization also depends on the structured gas metallicity $Z_g^{GMC}$. Metallicity in the dense phase is slightly higher than in the diffuse medium. 

\subsection{The attenuated stellar spectra}

A galaxy contains up to three distinct stellar populations associated with three different geometrical distributions. We build up the three spectral energy distributions (SED) associated with each stellar population:
\begin{itemize}
	\item{the young stellar population, in the disk: $(\lambda I_{\lambda})_{GMC}$,}
	\item{the old stellar population in the disk: $(\lambda I_{\lambda})_{ISM}$,}
	\item{the old stellar population in the central bulge: $(\lambda I_{\lambda})_{SPH}$.}
\end{itemize}

The overall extinguished stellar spectrum is then given by
\begin{equation}
	\begin{split}
		(\lambda I_{\lambda}) & = \Phi_{\lambda}^{ISM}\Phi_{\lambda}^{GMC}(\lambda I_{\lambda})_{GMC} \\ 
		 						& + \Phi_{\lambda}^{ISM}(\lambda I_{\lambda})_{ISM} \\
		 						& + \Phi_{\lambda}^{SPH}(\lambda I_{\lambda})_{SPH}	
	\end{split}
	\label{eq:extinguished_stellar_spectrum}
\end{equation}

\subsection{Infrared energy budget}

The bolometric luminosity produced by stars is:
\begin{equation}
	L_{\star} = \int(I_{\lambda})_{\star}d\lambda~~[L_{\odot}]   
	\label{eq:stellar_bolometric_luminosity}
\end{equation}
where $(I_{\lambda})_{\star}$ is the non-extinguished stellar SED (Sect. \ref{sec:stellar_spectra}).\\
 
By assuming that dust in the diffuse ISM and in GMC absorbs a fraction of this radiation and re-emits it the IR range, the energy balance leads to a total IR luminosity given by:
\begin{equation}
	L_{IR} = L_{\star} - \int(I_{\lambda})d\lambda~~[L_{\odot}]   \,,
	\label{eq:total_Lir}
\end{equation} 
where $(I_{\lambda})$ is the extinguished stellar SED build throught Eq. \ref{eq:extinguished_stellar_spectrum}.

\subsection{Effective dust attenuation}

The  attenuation is a combination of the three distinct attenuations associated with the distinct stellar populations. For simplicity, and to allow a comparison with observations, we define the effective attenuation:
\begin{equation}
	\Phi_{\lambda}^{eff} = \dfrac{(\lambda I_{\lambda})}{(\lambda I_{\lambda})_{\star}}~~~ \mathrm{and} \,\,\, A_{\lambda}^{eff} = -2.5log_{10}\Phi_{\lambda}^{eff} \,.
	\label{eq:effective_attenuation}
\end{equation}

\begin{figure}[t!]
	\begin{center}
		\includegraphics[scale=0.7]{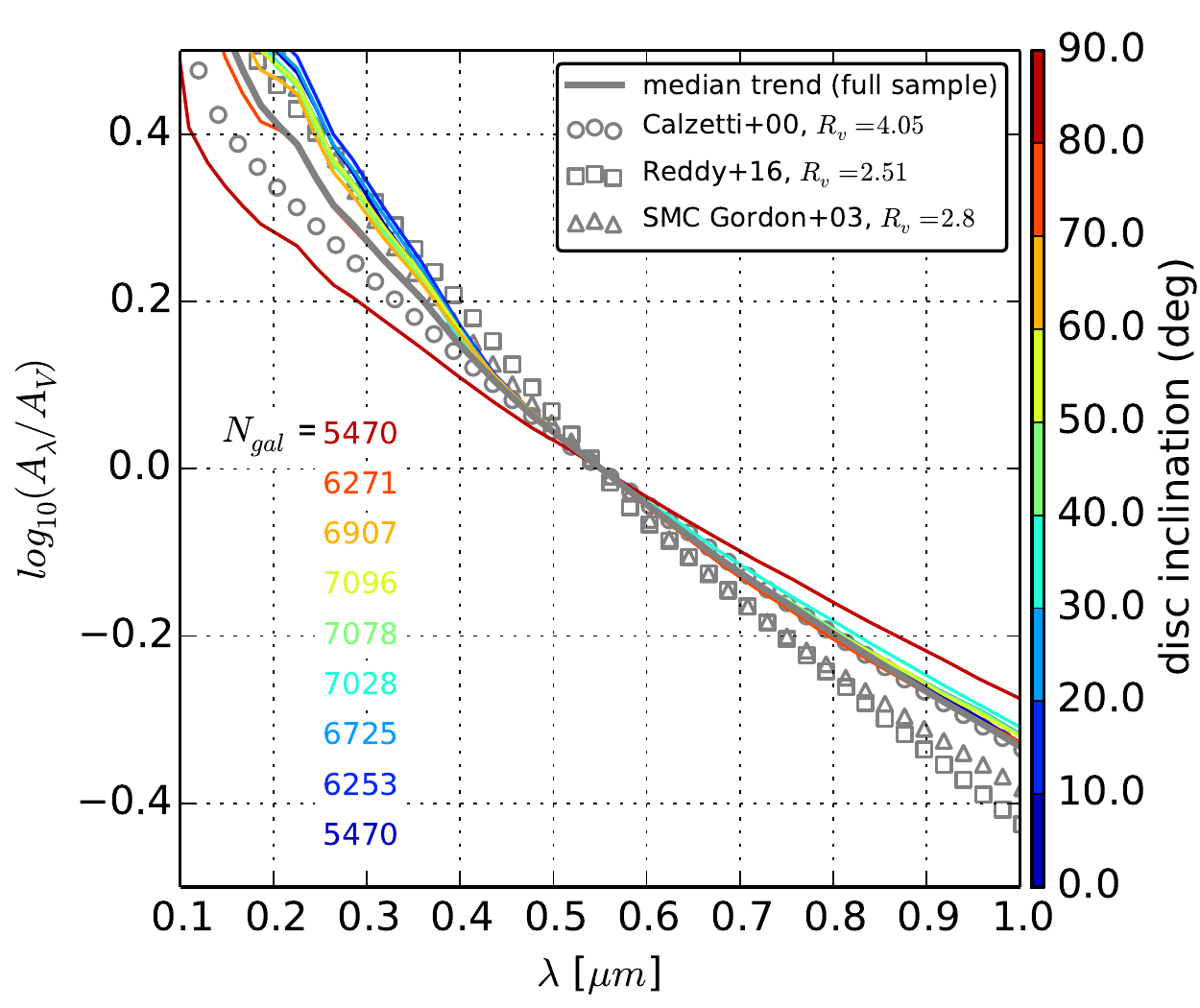}
  		\caption{Median effective attenuation curves as functions of the disk inclination for a random star-forming galaxy sample extracted between z = 4 and z = 1.5. The galaxy sample has been divided in 9 bins of disk inclination, from face-on ($0^{\circ}$) to edge-on disks ($90^{\circ}$). The color code indicates different disk inclinations. The median trend of the full sample is plotted with the solid gray line. We compare this trend with a set of standard attenuation curves, \cite{Calzetti_2000}: circles, \cite{Reddy_2016}: squares and \cite{Gordon_2003} (SMC): triangles.}
  		\label{fig:effective_attenuation_inclination_dependency}
	\end{center}
\end{figure}

Once dust/stars geometry and dust properties are fixed, the FUV depth (used as normalization) and the disk inclination are the two last key parameters for the effective attenuation. Fig. \ref{fig:effective_attenuation_inclination_dependency} shows the impact of disk inclination. As expected \citep[e.g.][]{Chevallard_2013}, face-on disks show a steeper attenuation curves than edge-on disks. Between these two extreme cases we observe a slow evolution of the slope of the attenuation curve. We compare these median attenuation curves to both \cite{Calzetti_2000} and \cite{Reddy_2016} laws. In the wavelength range $\lambda\in [0.5 - 1]\,\mu$m  our median attenuation curves are in excellent agreement with \cite{Calzetti_2000} measurements. Our median attenuation curves corresponding to the highest inclinations are above that of \cite{Calzetti_2000}. For $\lambda\in [0.1 - 0.5]\,\mu$m and for the majority of disk inclinations, our median attenuation curves are steeper than the \cite{Calzetti_2000} law and closer to those measured by \cite{Reddy_2016}. However, our median attenuation curves are shallower than the attenuation law measured by \cite{Reddy_2016}. The sample of galaxies used to explore the evolution of the effective attenuation curve are randomly selected between z = 4.0 and z = 1.0 and for stellar mass $M_{\star} > 10^{7}\Msun$. It is important to note that for this large stellar mass window, our results do not vary with the redshift. 

\begin{figure}[t!]
	\begin{center}
		\includegraphics[scale=0.7]{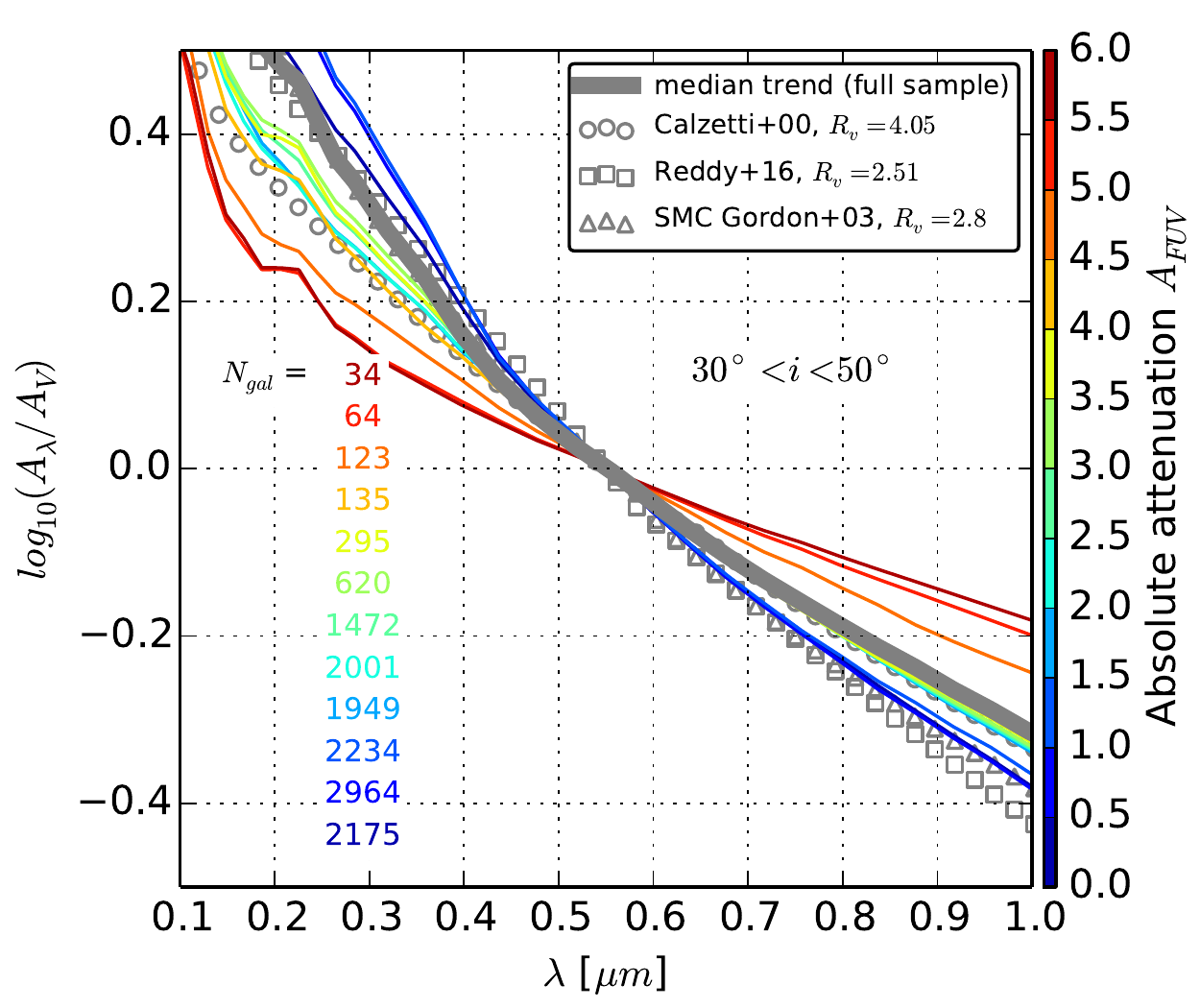}
  		\caption{Variation of the median effective attenuation curves with $A_{FUV}$ (color coded). We only select here galaxies with a disk inclination between $30^{\circ}$ and $50^{\circ}$. The median trend of the full sample is plotted with the solid gray line. We compare this trend to a set of standard attenuation curves, \cite{Calzetti_2000}: circles, \cite{Reddy_2016}: squares and \cite{Gordon_2003} (SMC): triangles.}
  		\label{fig:effective_attenuation_AFUV_dependency}
	\end{center}
\end{figure}

In a given disk inclination bin, the variation of the optical depth in FUV band leads to a large scatter around the median trend. This scatter can be larger than the one resulting from the variation of inclination. To illustrate this scatter we select galaxies with inclinations between $30^{\circ}$ and $50^{\circ}$. In this sub-sample, we show in Fig.\ref{fig:effective_attenuation_AFUV_dependency} the trend of the attenuation curve as a function of the absolute attenuation in the FUV band ($A_{FUV}$). As expected, a lower $A_{FUV}$ is associated with a steeper attenuation curve. In the wavelength range $\lambda\in [0.1 - 0.5]\,\mu$m, the median trend of this sub-sample of galaxies lies between the \cite{Calzetti_2000} and the \cite{Reddy_2016} attenuation curves. Galaxies experiencing a very low FUV attenuation can have a steeper attenuation curve than the measurements of \cite{Reddy_2016}. In contrast, galaxies with a strong attenuation in FUV band show a shallower attenuation curve than \cite{Calzetti_2000} law. According to these trends, the absolute extinction in FUV band seems to be the main driver of the extinction even if it is difficult to extract a unique main contributor driving the attenuation curve slope.

The observed scatter of the distribution of our effective attenuation curves is fully consistent with recent models implemented in hydrodynamic simulations, especially the recent study performed at z = 5.0 by \cite{Cullen_2017} (their Fig. 12) in the context of the First Billion Years projects.

%
%

\section{Luminosity functions}
\label{sec:Luminosity_functions}

\begin{figure*}[t!]
	\begin{center}
	  	\includegraphics[scale=0.48]{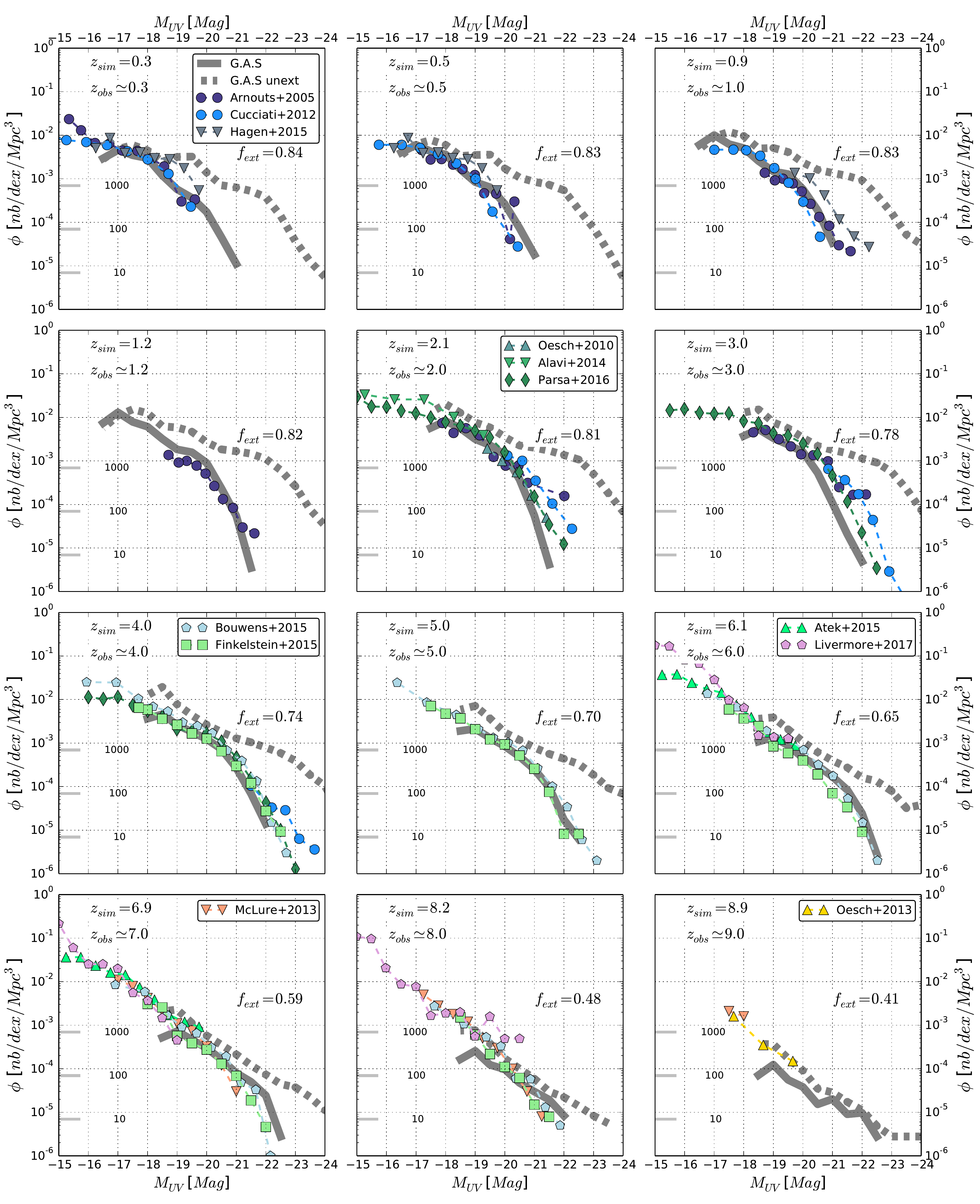}
  	 	\caption{UV luminosity functions. The different panels correspond to different redshift ranges. Predictions of our model are compared to a set of observations: \cite{Arnouts_2005}: dark-blue circles, \cite{Oesch_2010}: dark-green up-triangles, \cite{Cucciati_2012}: blue circles, \cite{McLure_2013}: orange down-triangles, \cite{Oesch_2013} yellow up-triangles, \cite{Alavi_2014}: green low-triangles, \cite{Parsa_2016}: dark-green diamonds, \cite{Hagen_2015}: grey low-triangles, \cite{Bouwens_2015} lightblue pentagons, \cite{Finkelstein_2015}: lightgreen squares, \cite{Atek_2015b}: light-green up-triangles and \cite{Livermore_2017}: purple pentagons. In each panel $f_{ext}$ indicates the fraction of the intrinsic UV emission of stars, which is attenuated by dust. The tick marks on the left side of each panel allow to link the comoving density to the effective number of galaxies taken into account in the luminosity function.}
  		\label{fig:UV_luminosity_functions}
	\end{center}  
\end{figure*} 

We now compare our predictions with some luminosity function measurements. We focus on the UV Luminosity Function (UVLF) and the IR Luminosity Functions (IRLF). Luminosity functions are built for stellar masses greater than $10^{7}M_{\odot}$. This mass threshold corresponds to our stellar mass resolution. Close and below this mass, our predictions are incomplete and suffer of numerical instabilities. Particularly, gas-phase metallicity cannot be computed and dust composition cannot be defined. According to the star formation histories and the evolution of the  stellar populations metallicity, this mass threshold corresponds to different UV magnitudes and IR luminosities limits and varies with z. In Figs. \ref{fig:UV_luminosity_functions} and \ref{fig:IR_luminosity_functions} we present luminosity function only above the turnover, which is specific of the incompleteness of the data.

\subsection{UV luminosity functions}

\subsubsection{The overall evolution}

Fig. \ref{fig:UV_luminosity_functions} shows the predictions for the UVLF at different redshifts. UV Magnitudes have been computed by convoluting the rest-frame spectrum of each galaxy with the GALEX \citep{Martin_2005a} FUV transfer function\footnote{Obviously, similar results are obtained by convoluting for example i, r, or u filter bands with the redshifted galaxy spectrum.}. We compare our predictions with a large set of observational data coming from \cite{Arnouts_2005}, \cite{Oesch_2010}, \cite{Cucciati_2012}, \cite{McLure_2013}, \cite{Oesch_2013}, \cite{Alavi_2014}, \cite{Parsa_2016}, \cite{Hagen_2015}, \cite{Bouwens_2015}, \cite{Finkelstein_2015}, \cite{Atek_2015b} and \cite{Livermore_2017}. 

For all redshifts, we show the attenuated and intrinsic luminosity function. At all redshifts, the attenuated UVLFs and measurements are in very good agreement. However, at z$\simeq$9.0 and at z = 3.0 our predictions are slightly below the observations, especially in the bright end part of the luminosity function at z = 3.0 and in the low-luminosity range at z$\simeq$9.0. At z$\simeq$9.0 the number of galaxies taken into account in the predicted UVLF is low ($<$500). At these very high redshifts, measurements are based on only few galaxies and may suffer from strong uncertainties. UVLFs are therefore affected by statistical noise. At $z\simeq$3.0, UVLFs are built with a larger number of objects: the discrepancy between observational measurements and prediction cannot be only explained by uncertainties. At $z\simeq3.0$ specifically, modelled galaxies seem to be affected by a too strong attenuation, especially in the high luminosity (and also mass) range. This trend is also confirmed by the behavior of the K-band luminosity function at this redshift (Cousin, in preparation). The IR luminosity function (Fig. \ref{fig:IR_luminosity_functions}) indicates that the IR energy absorbed by dust in this high luminosity regime is in agreement with observations. In addition, the stellar mass function presented in paper I also shows a good agreement with measurements. The discrepancy noted at $z\simeq3.0$ cannot be therefore currently explained. 

At all redshifts and for the range of magnitudes considered here, the shape of the intrinsic UVLF is compatible with a power law. At highest luminosities a break appears. It is only due to the exponential decrease of the number of galaxies present in the volume. On the contrary, the shape of the extinguished UVLF evolves with both the redshift and the UV luminosity. At high redshifts ($z\ge 6.0$), the shape of the extinguished UVLF is still compatible with a power law but with a steeper slope than the un-extinguished UVLF. For redshifts z$<$6.0, the extinguished UVLF clearly shows two different shapes. In the low luminosity range $M_{UV} > -20$, the shape is compatible with a power law. However, in the high luminosity range ($M_{UV} < -20$), we find an exponential decrease.

\subsubsection{How many energy is absorbed by dust ?}

By comparing the attenuated (ext) and the intrinsic (int) UV luminosities of all the galaxies produced by our model at a given redshift, it is possible to compute the fraction of UV radiation absorbed by dust. At a given redshift, we sum separately the extinguished and the intrinsic UV luminosities over all galaxies. We define the UV extinguished fraction as the ratio of these two quantities:
\begin{equation}
	\displaystyle f_{ext} = 1.0 - \dfrac{\sum L_{FUV}^{ext}}{\sum L_{FUV}^{int}}
	\label{eq:extinguished_UV_luminosity}
\end{equation}

This fraction is indicated in each panel of Fig.\,\ref{fig:UV_luminosity_functions}. Our model predicts an UV extinguished fraction of 40\% at z$\simeq$9.0. At this redshift, only the most massive galaxies ($M_{\star} > 10^{10.5}\Msun$) are affected by dust extinction. For the galaxies with lower mass, the average metallicity is too low. The fraction of UV radiation absorbed by dust progressively increases when redshift decreases. At z$\simeq$8.2 our model predicts that roughly half of the UV radiation produced by young stars is absorbed by dust. This fraction reaches 60\%, 70\% and 80\% at z $\simeq$ 7.0, 5.0 and 3.0. Obviously this progressive increase of the dust attenuation is linked to the progressive increase of the gas-phase metallicity in galaxies. At lower redshift (z$\le$ 2.0), the fraction of UV radiation absorbed by dust still increase but much more slowly. At z$\simeq$0.3 the fraction of UV radiation absorbed by dust reaches a value of $\simeq 86$\%.

By integrating a compilation of UV and IR luminosity functions, \citet{Burgarella_2013} found a similar ratio of 85$\pm$10\,\% and 84$\pm$13\,\% at z=2.2 and z=3.15, respectively. These results agrees at less than 1$\sigma$ with our model, which predicts 82 and 79\,\%, respectively. At higher redshifts, only weak constraints are available in the far-infrared \citep{Madau_2014}, since the limited depth of the instruments does not allow us to constrain accurately the position of the knee of the luminosity function. \citet{Burgarella_2013} used the attenuations estimated by \citet{Bouwens_2009} to put constraints at higher redshifts and found that the equality between UV and IR (f$_{\rm ext}$ = 50\,\%) is reached at z$\simeq$6. Several teams also tried using various modeling approaches to estimate the evolution of the ratio between obscured and unobscured star formation density. A similar value is found by the empirical model of \citet{Bethermin_2017} and in the measurement of the obscured star formation rate density from the CIB anisotropies from \citet{Maniyar_2018}. In contrast, it disagrees with \citet{Koprowski_2017} who predicts a transition around z$\simeq$3 based on their empirical modeling approach.

\subsection{IR luminosity functions}
\label{sec:IR_luminosity_functions}

\begin{figure*}[t!]
	\begin{center}
		\includegraphics[scale=0.48]{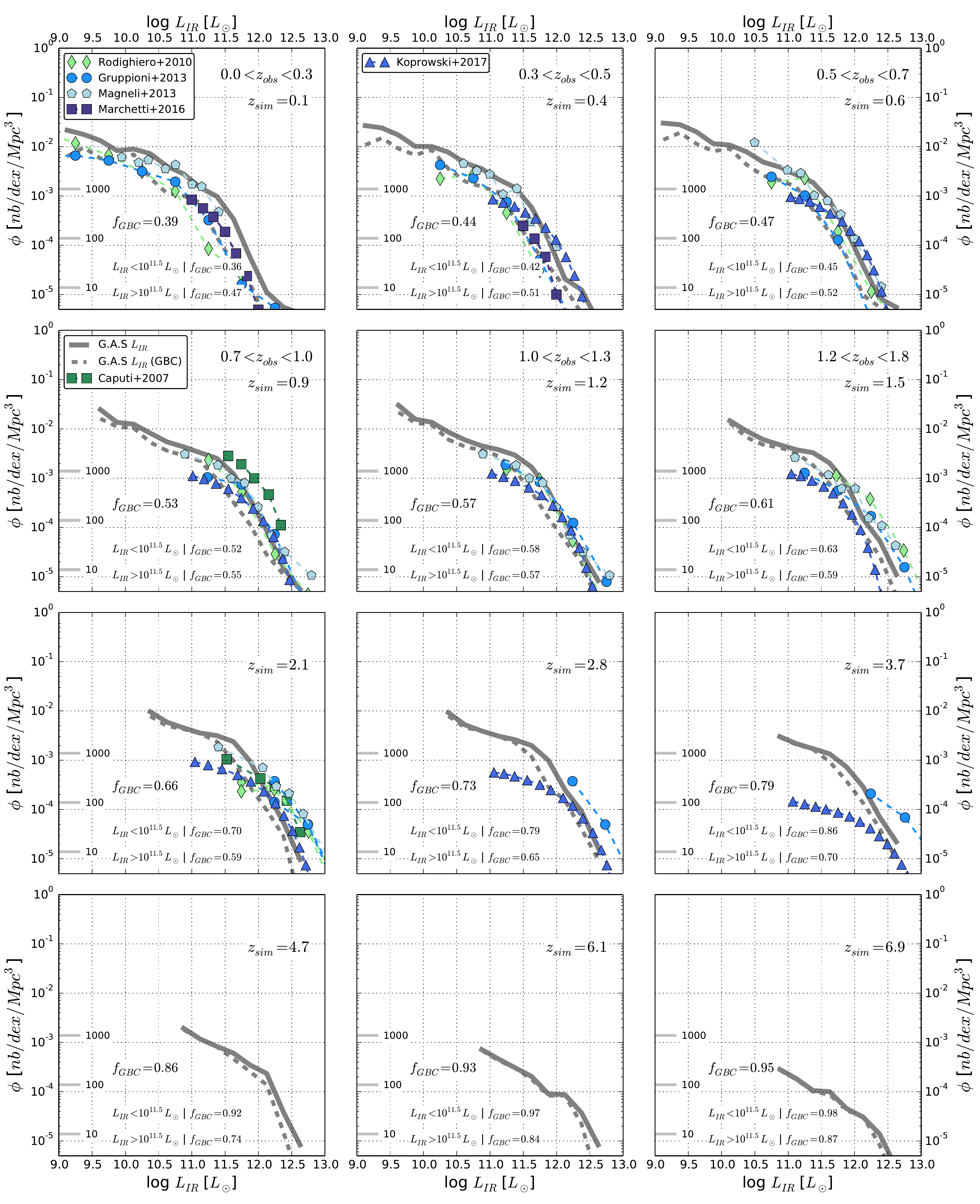}
  		\caption{IR luminosity functions. The different panels are associated with different redshift ranges. \GAS\ model predictions are only plotted above the stellar mass completeness limit ($10^{7}M_{\odot}$). We compare our results with measurements done by \cite{Caputi_2007}: green squares, \cite{Rodighiero_2010}: lightgreen diamonds, \cite{Gruppioni_2013}: blue circles, \cite{Magnelli_2013}: lightblue pentagons, \cite{Marchetti_2016}: purple squares and \cite{Koprowski_2017}: dark blue triangles. The solid grey line gives the total IR luminosity function (including diffuse disk ISM, GMCs, and bulge). The gray dashed line gives the IR luminosity function that takes into account only the GMC environment and therefore only the young stars contribution. The tick marks on the left side of each panel allow to link the comoving density to the effective number of galaxies taken into account in the luminosity function.}
  		\label{fig:IR_luminosity_functions}
	\end{center}
\end{figure*}

\subsubsection{The evolution of the total IR luminosity function}

Fig.\,\ref{fig:IR_luminosity_functions} shows our predictions for the total IR luminosity function at different redshifts. We compare our results with measurements coming from \cite{Caputi_2007}, \cite{Rodighiero_2010}, \cite{Gruppioni_2013}, \cite{Magnelli_2013}, \cite{Marchetti_2016} and \cite{Koprowski_2017}. 

Our predictions are in general in very good agreement with observations. At all redshifts presented here, the total IR luminosity function predicted by our \GAS\ model is fully compatible with a double power-law. A knee between the two regimes (low and high luminosities) appears around $L_{IR}\simeq10^{12}L_{\odot}$ at z$\simeq$6.0. The break progressively shift to lower luminosity. At $z\le1.0$ the knee appears around $L_{IR}\simeq10^{11.5}L_{\odot}$. 

For the redshifts z$\simeq$2.8 and $\simeq$3.7, even if the bright ends are in good agreement with the recent \cite{Koprowski_2017} estimates, \GAS\ seems to over-predict the density of IR galaxies for $L_{IR} < 10^{12}L_{\odot}$. However we also note strong tensions between different measurements at these high redshifts. 

Even if measurements are not available at z $>$ 4.0, we show the \GAS\ predictions at z = 4.7, 6.1 and 6.9 in the three lower panels of Fig.\,\ref{fig:IR_luminosity_functions}. These IRLF highlight that IR luminosities higher than $10^{12}L_{\odot}$ can already be reached by galaxies at z$\simeq$7. Metal production and dust assembly is therefore already very active at this epoch. This prediction is in agreement with hydrodynamic simulations, other SAM taking into account dust obscuration \cite[e.g.][]{Clay_2015} and also recent observations of high redshift (z$>$6) heavily-obscured galaxies \citep[e.g.][]{Bowler_2018, Watson_2015}.     

\subsubsection{Contributions of the different environments, Is there an excess of cold dust at low redshift ?}

In addition to the total IR luminosity function, we also plot in each panel of the Fig. \ref{fig:IR_luminosity_functions} the luminosity function build only with the IR luminosity coming from GMC environments (young stars). The fraction $f_{GMC}$ of the IR luminosity coming from GMC environments is also indicated in each panel.  

At high redshift ($z > 3.5$) this fraction reaches very high values ($>$80, 95\%). The large majority of the IR light comes from dense strongly attenuated GMCs. Between $z=1.5$ and $z=3.5$ this fraction is still larger than 60\%. For $z<1.5$ the contribution of GMC progressively decreases and reaches $\simeq$ 53\% at $z = 0.9$. The contribution of the oldest stars evolving in the diffuse ISM appears progressively. However, the majority of the IR luminosity always comes from GMCs illuminated by young stars.

At $z\le0.6$ the fraction of IR luminosity coming from the GMC environment drops below 50\% and even reaches 39\% at z$\simeq$0.1. The contribution of old stars is therefore slightly dominant. In parallel, at $z = 0.4$ and $z=0.1$, our \GAS\ model generates a slight over-density of IR galaxies. The IR emission is slightly too high for the full IR luminosity range explored. However, if we focus only onto the GMC contribution the agreement at these low redshift is very good. The IR emission coming from the cold dust component hosted by the diffuse ISM is probably too high. This trend is confirmed in the \GAS III paper. IR number counts and the comparison of the IR SEDs with recent measurements reveal an excess of IR light emitted by the cold dust component, mainly in the massive local ($z<0.5$) galaxy population.

\subsubsection{The impact of the latest disruption episode on massive IR bright galaxies}

At all redshifts explored here, the gap existing between the total and the GMC-only IR luminosity functions seems indicated that the contribution of the GMC to the total IR luminosity seems lower at the bright-end than at the faint-end of the IR luminosity function. To explore this trend we compute the contribution of GMCs for both the faint IR-galaxies ($L_{IR} < 10^{11.5}L_{\odot}$) and the bright IR-galaxies ($L_{IR} > 10^{11.5}L_{\odot}$). These median fractions are indicated in the bottom of each panel of Fig. \ref{fig:IR_luminosity_functions}. 

At all redshifts shown here, the contribution of the GMCs ($f_{GMC}$) is found to be larger in luminous IR galaxies ($L_{IR}>10^{11.5}L_{\odot}$) than in faint IR galaxies ($L_{IR}<10^{11.5}L_{\odot}$). This trend is especially true at high redshifts. At z $\le$ 1.5, the contributions of the two galaxy populations are closer and vary between 60\% and 40\%.

This trend is explained by the progressive evolution with the stellar mass (IR luminosity) of the dense/fragmented gas fraction. In the star-forming galaxy sample, massive/IR-luminous galaxies ($M_{\star} \ge 10^{10.5}M_{\odot}$, $L_{IR} \ge 10^{10.5}L_{\odot}$) the gas content is less structured/fragmented than in less massive and less IR-luminous galaxies. In massive galaxies, we observe a progressive decrease of the gas accretion. This leads to a decrease of the star formation activity. However, SN feedback still disrupts the dense/fragmented gas after the last star formation episode (\GAS\ I paper). The impact of the latest disruption episode impacting the GMCs leads to a more diffuse ISM and therefore to an attenuation slightly lower in massive (IR-luminous) galaxies than in the less massive ones.

%
%

\section{Dust attenuation}
\label{sec:dust_attenuation_IRX}

The IRX can be used to measure the attenuation in FUV $A_{FUV}$. Many efforts have been made to link IRX (and therefore $A_{FUV}$ ) to galaxy properties like the slope of the UV spectrum, the stellar mass or the star formation rate \citep[e.g.][]{Meurer_1999, Buat_1999, Gordon_2000, Bell_2002, Kong_2004, Calzetti_2005, Seibert_2005, Cortese_2006, Boissier_2007, Salim_2007, Salim_2009, Treyer_2007, Boquien_2009, Buat_2010, Takeuchi_2010}. 

In this section we explore the variations of IRX with galaxy properties such as gas-phase metallicity and stellar population (age and mass). For the IRX-$\beta$ relation we also analyze its dependency with disk inclination and slope of the attenuation curve.

Our analysis is based on a star-forming galaxies sample. We only keep galaxies with a mass higher than $10^{8}M_{\odot}$. A star-forming galaxy is defined following the \textit{Main Sequence} SFR--$M_{\star}$ relation calibrated by \cite{Schreiber_2015}. We consider that a galaxy is star forming if its specific star formation rate sSFR is greater\footnote{We checked that our results are not significantly affected if we use instead a threshold of 1/3 or 1/5.} than $1/4\times (SFR_{MS}/M_{\star})$. 

\subsection{The IRX -- A$_{FUV}$ relation}

In star forming galaxies, the link between IRX and $A_{FUV}$ has been explored by \cite{Buat_2005}, \cite{Cortese_2008} or \cite{Boquien_2012} and also by using radiative transfer models \citep[e.g.][]{Witt_2006}.

\begin{figure}[t!]
	\begin{center}
		\includegraphics[scale=0.7]{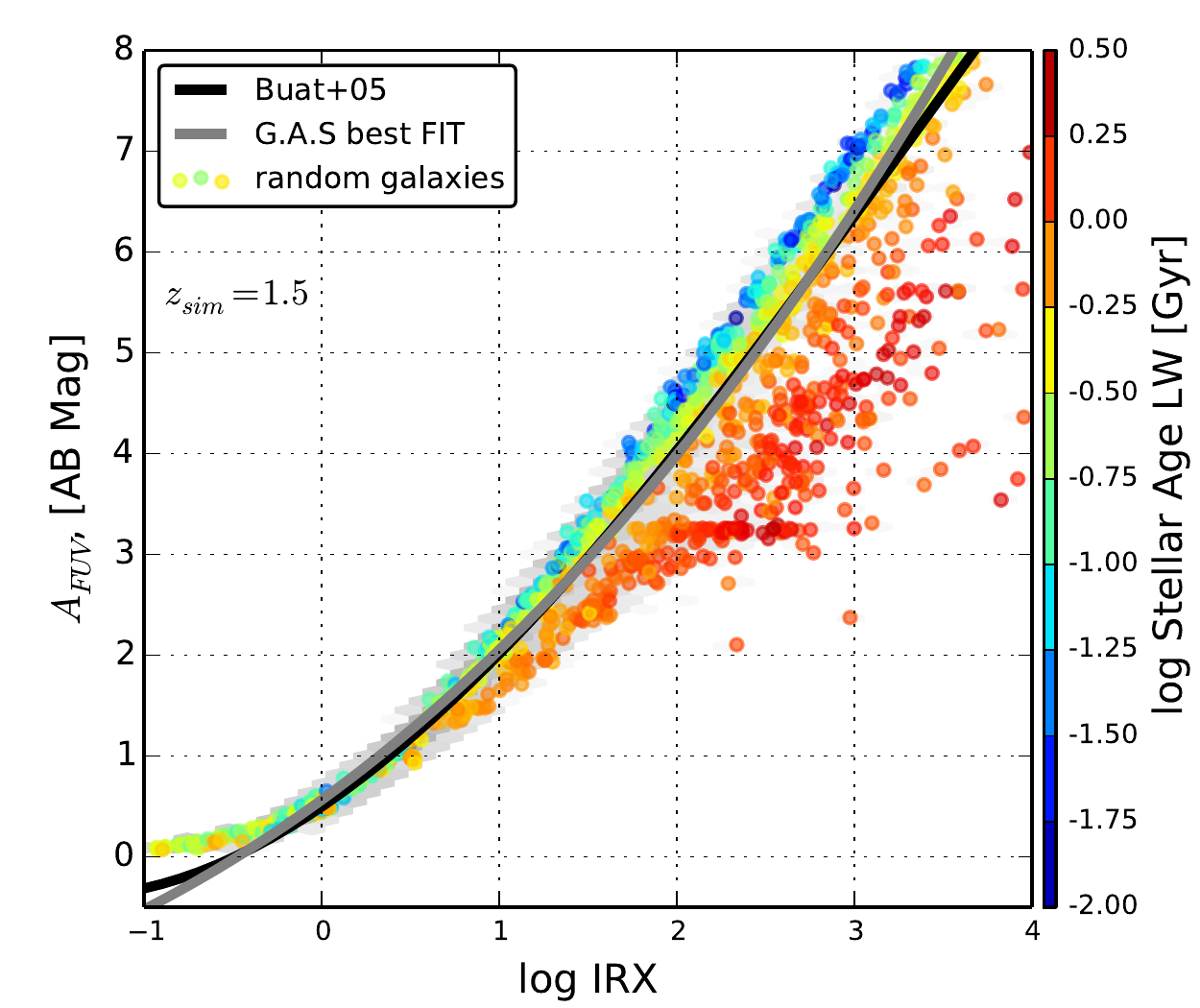}
  		\caption{IRX-A$_{FUV}$ relation derived from a star forming galaxies sample extracted at z = 1.5. The gray solid line marks our best fit. For comparison we also plot the \cite{Buat_2005} best fit (black solid line). Colored points show a sub sample that is uniformly selected in the IRX-A$_{FUV}$ plan. The color scale shows the luminosity-weighted age of the stellar population hosted by the galaxy.}
  		\label{fig:IRX_Afuv_age_dependency}
	\end{center}
\end{figure}

Fig.\,\ref{fig:IRX_Afuv_age_dependency} shows the IRX-A$_{FUV}$ relation based on a star forming galaxies sample extracted at z = 1.5. Our predictions agrees with the relation from \cite{Buat_2005}. 

Even if the large majority of galaxies lies around the parametric relation, we see from Fig.\,\ref{fig:IRX_Afuv_age_dependency} that a scatter progressively appears at high IRX and high A$_{FUV}$. As highlighted on Fig.\, \ref{fig:IRX_Afuv_age_dependency} this scatter is linked to the age of the stellar population hosted by galaxies. At a given IRX, a lower A$_{FUV}$ is associated with an older stellar population. Similarly, at a given A$_{FUV}$, larger IRXs also corresponds to older stellar populations. We clearly see that the scatter of the relation increases with IRX. The variation with the luminosity-weighted stellar population age observed in our {\tt G.A.S} model is in agreement with the analysis done by \cite{Cortese_2008}. They conclude that, at a given IRX, active star forming galaxies (with young stars) are more attenuated than passive galaxies (with old stars). In Fig.\, \ref{fig:IRX_Afuv_age_dependency}, our results are obtained at z=1.5, but similar trends and behaviors can also be extracted at other redshifts. We do not observe any variation of the IRX-A$_{FUV}$ relation with the redshift.

\subsection{The IRX -- gas metallicity relation}
\label{sec:IRX_Metallicity}

\begin{figure}[t]
	\begin{center}
		\includegraphics[scale=0.72]{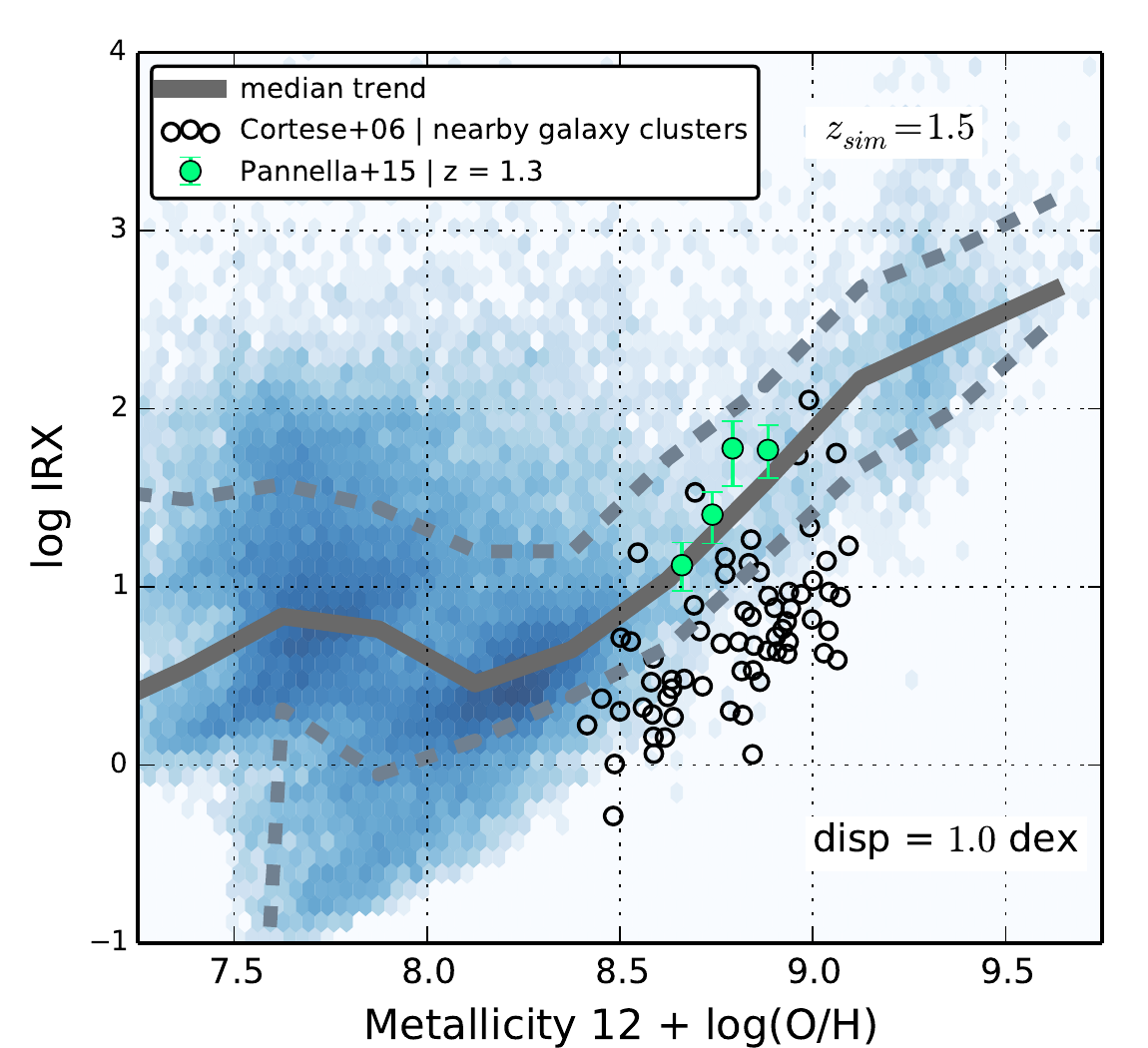}
  		\caption{IRX--$Z_{g}$ relation for star forming main sequence galaxies extracted at z = 1.5. The gas-phase metallicity is computed as the Oxygen over Hydrogen abundance: $Z_g = 12 + log(O/H)$. We focus here on the dense/fragmented gas phase metallicity. The blue density shows the complete star-forming galaxy sample. The solid gray line shows the median trend of this sample. Dashed gray lines indicate the 15\% and 85\% quantiles of the distribution. We compare the {\tt G.A.S.} model with observational measurements coming from \cite{Pannella_2015} and \cite{Cortese_2006}.}
  		\label{fig:IRX_Metallicity}
	\end{center}
\end{figure}

The production of dust in the ISM is closely linked to the metal enrichment of the gas phase and a correlation between IRX and gas phase metallicity is expected. Such a correlation has been found in nearby galaxies by e.g. \cite{Boquien_2009}. In their sample, galaxies with higher metallicities have higher IRX.

Fig.\,\ref{fig:IRX_Metallicity} shows the IRX--$Z_g$ relation built from our z=1.5 star forming galaxy sample. As discuss previously, the effective attenuation is mainly produced by the GMC environment associated with the dense/fragmented gas phase. The average gas metallicity used in Fig.\,\ref{fig:IRX_Metallicity} is therefore computed in the dense/fragmented gas phase according to Eq. \ref{eq:gas_metallicity}. 

In Fig.\,\ref{fig:IRX_Metallicity}, the majority of our star forming galaxies lie at $Z_g < 8.5$. In this low metallicity regime, the dispersion is high (close to 2\,dex) and it is  difficult to extract any trend between the IRX and the gas-phase metallicity.

However, at high metallicity, i.e. $Z_g > 8.5$ we observe a clear correlation with the IRX increasing with the average gas phase metallicity. In this domain, the median trend is in good agreement with the measurements performed by \cite{Pannella_2015}. However, the data points obtained from star-forming galaxies evolving in nearby clusters by \cite{Cortese_2006} mainly lie below our median trend. At a given metallicity, local star forming galaxies are therefore less attenuated. This behavior could be associated with a less structured and less dense ISM in low redshift galaxies. As discuss previously (Sect. \ref{sec:IR_luminosity_functions}), when the redshift decreases, we observed in our simulated galaxies a decrease of the structured gas fraction. In parallel we can also note an increase of the characteristic size (exponential radius) of disks. These trends are especially marked in massive galaxies (i.e with $Z_g > 8.0$). Due to these two conjugated trends, the average gas density and the effective attenuation are therefore probably lower in low redshift galaxies ($z<1.0$) than at higher redshifts (here z = 1.5).

\subsection{The IRX--M$_{\star}$ relation}

\begin{figure}[t]
	\begin{center}
		\includegraphics[scale=0.7]{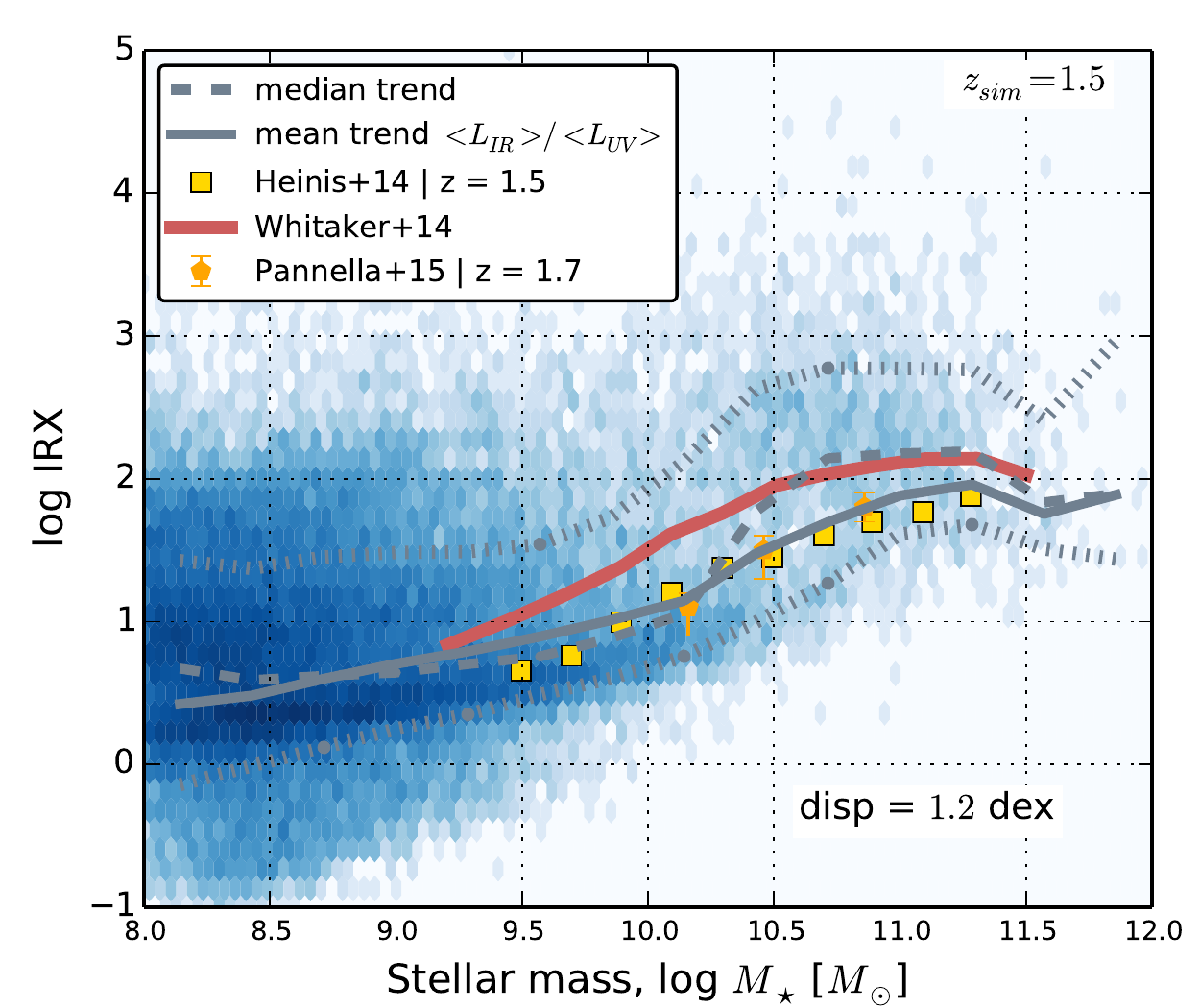}
  		\caption{IRX--$M_{\star}$ relation for star forming main sequence galaxies extracted at z = 1.5. The blue density shows the complete sample. The dashed gray line shows the median relation. The solid gray line shows the mean relation computed using IRX = $<L_{IR}>/<L_{UV}>$, to be consistent with what is measured by stacking. Around the median relation, dotted gray lines indicates the 15\% and 85\% quantiles of the distribution. We compare our predictions with observational measurements from \cite{Pannella_2015} (orange pentagons), \cite{Heinis_2014} (yellow squares) and \cite{Whitaker_2014} (solid red line).}
  		\label{fig:IRX_Mstars}
	\end{center}
\end{figure}

Our median IRX--$M_{\star}$ relation is plotted on Fig.\,\ref{fig:IRX_Mstars}. It agrees well with observational measurements \citep[e.g.][]{Heinis_2014, Whitaker_2014,Pannella_2015}. As observed, our model predicts a broad correlation between the stellar mass and the dust attenuation. In this context, the stellar mass has to be seen as a tracer of the galaxy activity, which is connected to the amount of metals produced by a galaxy since its formation \citep[e.g.][]{Cousin_2016,Tremonti_2004,Andrews_2013}. Through the stellar mass, we roughly trace the dust enrichment of the ISM.

It is often difficult to get both UV and IR luminosities for individual objects, especially at high redshift. The IRX is therefore mainly measured using stacking techniques \citep[e.g.][]{Bethermin_2012b, Hilton_2012}: for a given stellar mass, IRX is defined as the ratio of the average IR luminosity to the average UV luminosity: $IRX = <L_{IR}>/<L_{UV}>$ (stacking estimator). We compute this estimator in our different stellar mass bins. As shown in Fig. \ref{fig:IRX_Mstars}, we do not observe strong differences between our median and this stacking estimator in the low mass range ($M_{\star} < 10^{10}M_{\odot}$). At larger stellar masses $M_{\star} > 10^{10.5}M_{\odot}$, our stacking estimator reaches lower IRX values than the median trend and therefore, in this high mass regime, values and shape of the stacking estimator are in better agreement with observational measurements performed by \cite{Heinis_2014} and \cite{Pannella_2015}. In parallel, in this high mass regime, our median estimator shows a better agreement with \cite{Whitaker_2014}.

Around the median trend, we also plot in Fig. \ref{fig:IRX_Mstars} the dispersion limits of the relation. These two limits correspond to the 15\% and 85\% percentiles of the distribution in a given stellar mass bin. For stellar masses $M_{\star} > 10^{9.5}M_{\odot}$, we observe a large scatter around the median relation of about 1.2\,dex in average (and therefore a average dispersion of 0.6\,dex). This dispersion is in good agreement with the largest error bars of the measurements which are estimated to be 0.4-0.7\,dex \citep{Heinis_2014,Pannella_2015}. We note from Fig.\,\ref{fig:IRX_Mstars} that the dispersion mainly comes from the high IRX domain. Our {\tt G.A.S.} model predicts IRX larger than $10^{3}$ for all the stellar mass domain explored. These very extinguished galaxies are probably lost by the UV-selected sample used by \citep{Heinis_2014}.

\subsubsection{Correlation with UV luminosity}

\begin{figure}[t!]
	\begin{center}
		\includegraphics[scale=0.7]{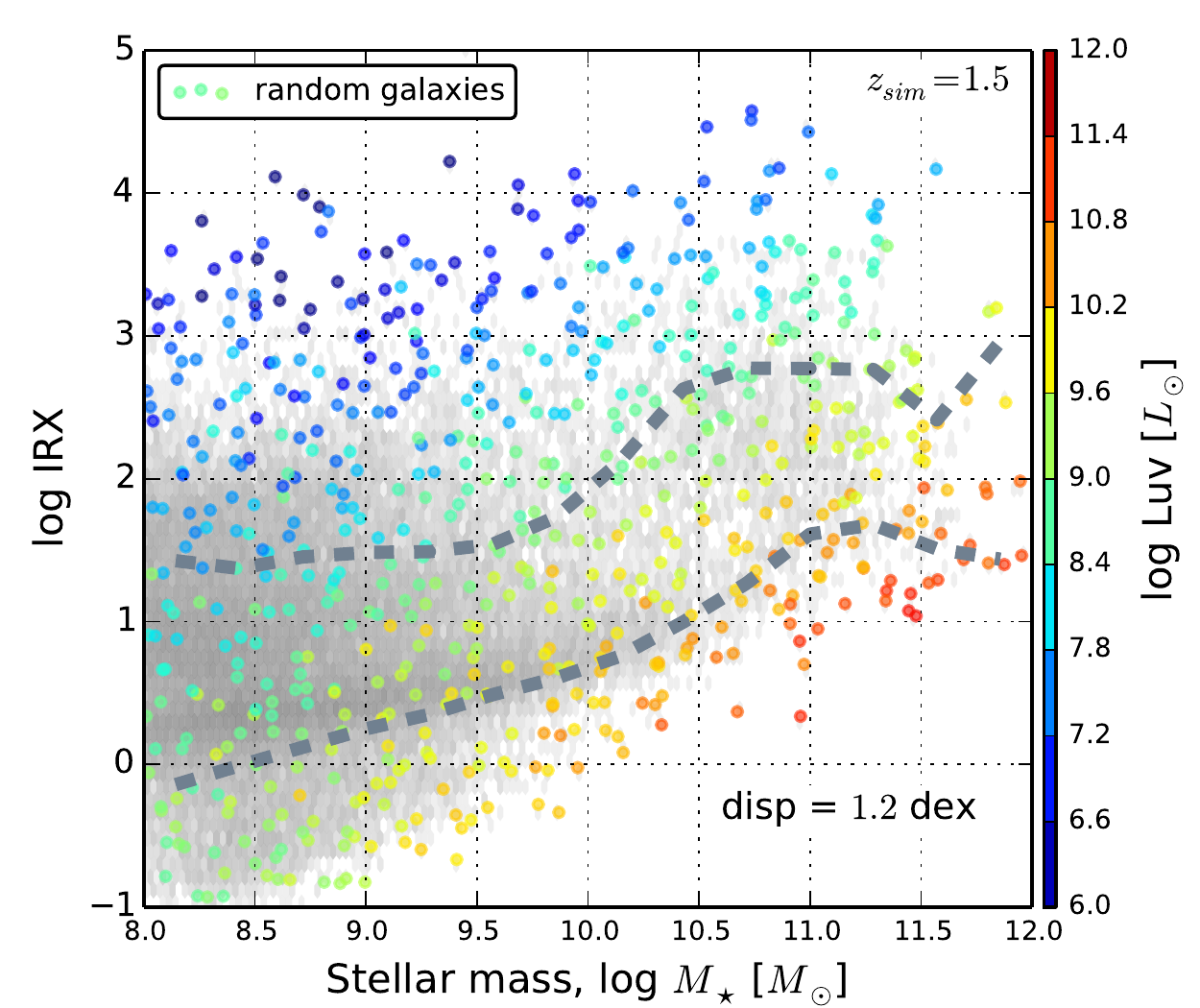}
  		\caption{IRX--$M_{\star}$ relation for main sequence galaxies at z = 1.5. We plot the 15\% and 85\% percentiles of the distribution (dashed grey lines). Colored points correspond to a random sample uniformly distributed in the IRX--$M_{\star}$ plan, with a color scale linked to the extinguished UV luminosity of the galaxies.}
  		\label{fig:IRX_Mstars_Luv_dependency}
	\end{center}
\end{figure}

In Fig.\,\ref{fig:IRX_Mstars_Luv_dependency}, we highlight the typical UV luminosity of galaxies depending on their position in the IRX--$M_{\star}$ diagram by selecting a sample uniformly in the IRX--$M_{\star}$ plan. Due to this homogeneous selection, the dispersion of these points is much larger than the scatter of the relation given above. This random sub-sample reveals a clear evolution of the extinguished UV luminosity with the IRX. At a given stellar mass, the observed UV luminosity decreases when the IRX increases in very good agreement with observational measurements of \cite{Heinis_2014} (see their Fig. 6, C1 and C2).

As mentioned previously, Fig.\,\ref{fig:IRX_Mstars_Luv_dependency} shows that our model predicts the existence of galaxies with a very strong attenuation (IRX $>10^{3}$). These galaxies are associated with $<10^{8}L_{\odot}$ observed UV luminosities, close from the lower limit of the \cite{Heinis_2014} UV-selected sample. These very UV-faint galaxies can therefore explain the \textit{large} scatter predicted by our model. To confirm this hypothesis, we limit our star-forming galaxies to galaxies with an extinguish UV luminosity higher than $L_{UV} > 10^{8.5}L_{\odot}$. With this sample, we do not observe any variation of our mean or median IRX--$M_{\star}$ relation, which are dominated by luminous UV galaxies. However the scatter of the relation falls to 0.9 and therefore the dispersion to 0.45\,dex which is fully compatible with dispersion measurements \citep{Heinis_2014}. 

\subsubsection{Evolution with redshift}

\begin{figure}[t!]
	\begin{center}
		\includegraphics[scale=0.66]{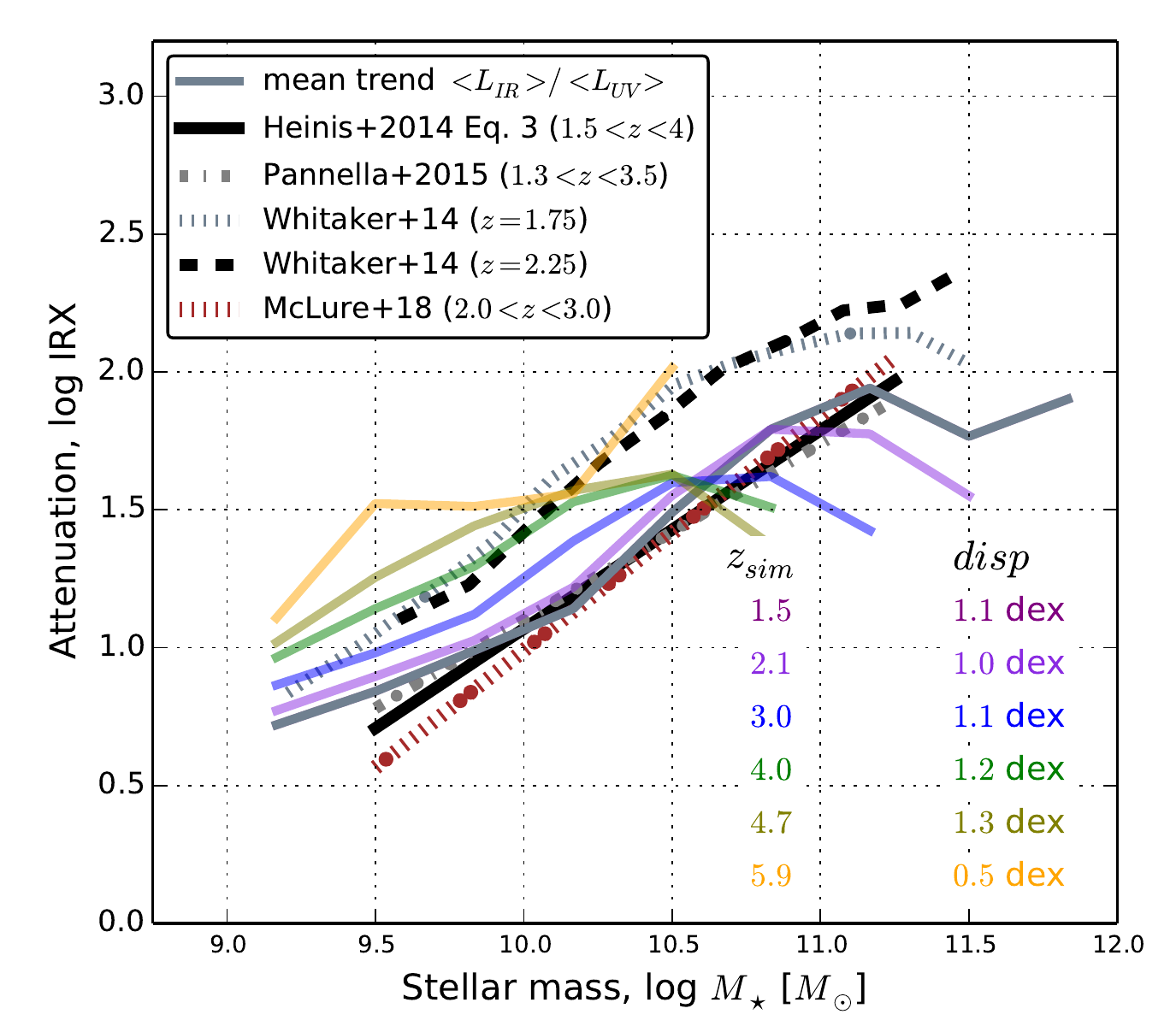}
  		\caption{Variation of the IRX--$M_{\star}$ relation for main sequence galaxies from z = 1.5 to z $\simeq$ 6.0. The colored solid lines show the mean (obtained using IRX = $<L_{IR}>/<L_{UV}>$) relation. Our predictions are compared to measurements coming from \cite{Whitaker_2014} at z = 1.75 and z = 2.25. We also add average relations from \cite{Heinis_2014}, \cite{Pannella_2015} and \cite{McLure_2018} which are available from $1.5<z<4.0$. In the bottom right corner are indicated the redshift of each sample and the associated dispersion.}
  		\label{fig:IRX_Mstars_Redshifts}
	\end{center}
\end{figure}

Fig.\,\ref{fig:IRX_Mstars_Redshifts} shows the evolution of the IRX--M$_{\star}$ relation with redshift (from z = 1.5 to z = 5.9) obtained using our stacking estimator. At all redshifts explored here, the scatter is similar and close to $\sim$1.2\,dex.

At stellar mass above $10^{10.5}M_{\odot}$, we observe a decrease of the IRX with the stellar mass. The mass threshold of this decrease shifts to lower mass when the redshift is increasing (from $\simeq10^{10.5}$ to $\simeq10^{9.5}M_{\odot}$ from z = 1.5 to z = 6, respectively). This decrease at high stellar mass is also observed by \cite{Whitaker_2014} with a trend similar to which is seen in our model. As previously (Sects. \ref{sec:IR_luminosity_functions}, \ref{sec:IRX_Metallicity}), we explain this turn-over with the evolution of the contribution of the GMC environment. Massive galaxies are less-structured and therefore show a lower attenuation than in less massive and more structured galaxies.

For M$_{\star}<10^{10.5}M_{\odot}$, at a given stellar mass, our model predicts an increase of the average IRX with the redshift. Even if we note some tension in the absolute IRX values measured\footnote{These studies use different observational data: \textit{Spitzer} for \cite{Whitaker_2014}, \textit{Herschel} for \cite{Heinis_2014} and \cite{Pannella_2015}} by \cite{Whitaker_2014} and \cite{Heinis_2014}, these two studies do not observe the increase predicted by our {\tt G.A.S.} model. For example, the average relations given by \cite{Heinis_2014} or \cite{Pannella_2015} are valid from z = 4.0 to z = 1.5. The discrepancy is mainly visible a $z>3.0$. At lower redshift the large dispersion ($\simeq1.1$\,dex) can explain the discrepancy. At $z>4.0$, IR and therefore IRX measurements are not currently available on a sufficiently large sample of galaxies. It is therefore difficult to put strong constraints on the amount of energy effectively absorbed by dust. At these high redshifts our UVLF agrees well with the measurements. To reduce the effective IRX and keep the UVLF similar, the overall attenuation has to be slightly reduced and the amount of young stars also reduced such as to keep a constant apparent UV emission. However, the amounts of stars currently predicted by \GAS at these redshifts agree well with the most recent measurements. In that context, it is difficult to explain the possible discrepancy.

\subsection{The IRX--$\beta$ relation}

\begin{figure}[t!]
	\begin{center}
		\includegraphics[scale=0.66]{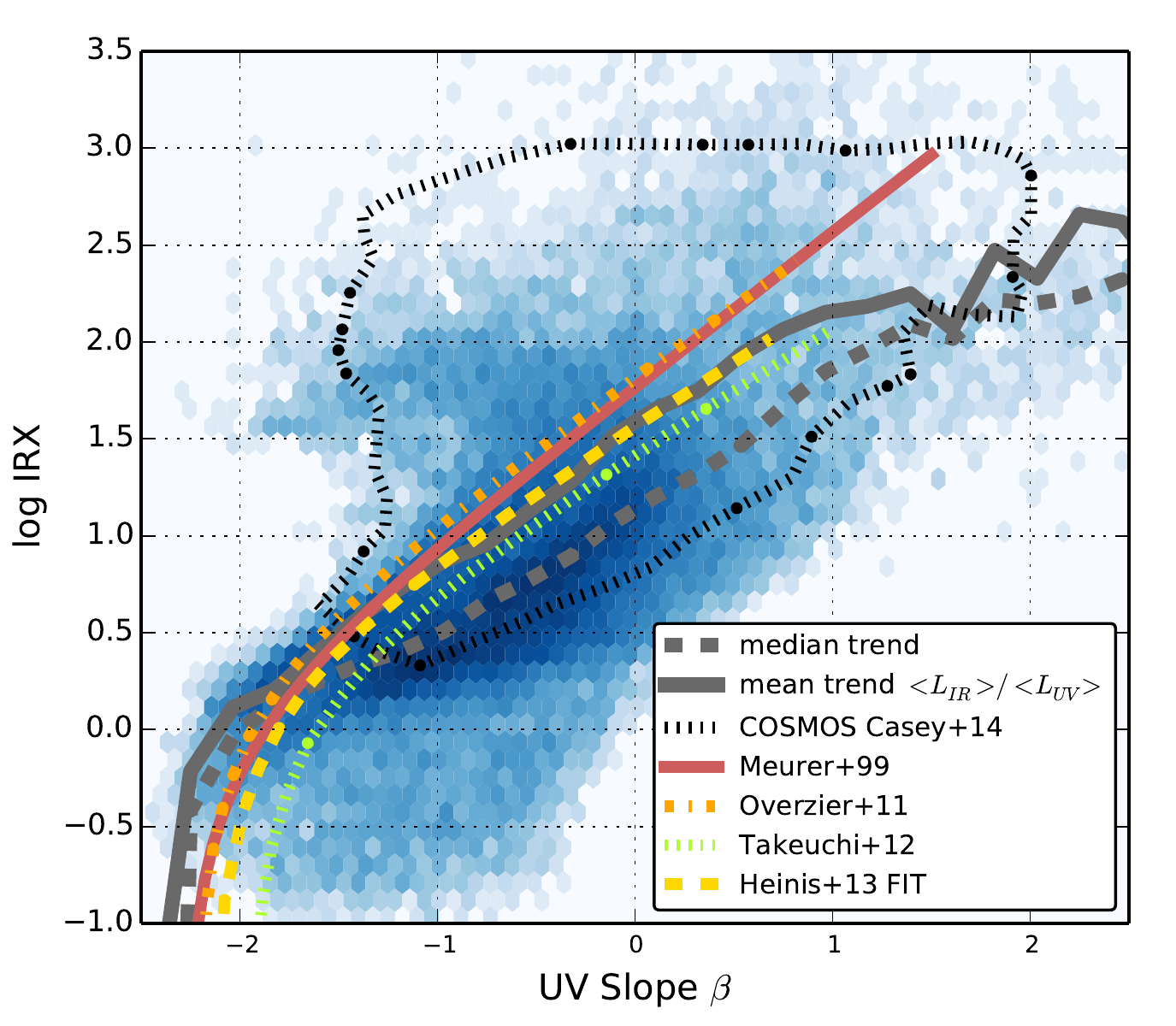}
  		\caption{IRX--$\beta$ relation for main sequence galaxies at $z = 1.5$. The median of the full sample (in blue) is shown with the dashed gray line. The IRX mean is shown with the solid gray line. Observations from \cite{Meurer_1999}, \cite{Overzier_2011}, \cite{Takeuchi_2012} and \cite{Heinis_2013} are shown in solid red, dot-dashed orange, dotted green and dashed yellow line, respectively. We also add contours coming from measurements performed in the COSMOS field by \cite{Casey_2014}.}
  		\label{fig:IRX_beta}
	\end{center}
\end{figure}

Fig.\,\ref{fig:IRX_beta} shows the IRX--$\beta$ relation based on our star forming galaxies sample extracted at z = 1.5. To compute the UV slope ($\beta$) we have fitted the rest-frame continuum of each galaxy with a power law expressed as $F(\lambda) \propto \lambda^{\beta}$. We used a set of five photometric points associated to filters 1, 3, 5, 7 and 9 listed in \cite{Calzetti_1994} (their Table 2). Similarly to what we did for the the IRX--M$_{\star}$ relation, we derived both the median trend of the IRX and the stacking estimator, i.e. the ratio of the mean IR luminosity to the mean UV luminosity in a given bin of $\beta$.

Our two estimators show similar behaviors, but the median is systematically lower than the mean. Our sample is in fact dominated by low-SFR galaxies ($<1\,M_{\odot}/yr$) and therefore low-luminosity galaxies. The median estimator simply reflects this distribution, while the stacking mean estimator is more sensitive to high SFRs (luminosity) values. Note that this bias also affect observational measurements.

Our mean relation agrees well with measurements done by \cite{Heinis_2013} and \cite{Takeuchi_2012}. It is important to note the large scatter of the distribution. Even if the majority of our simulated galaxies are located between $\beta\in$ [-1.5, 0.0] and log$_{10}$IRX$\in$ [0.0, 2.0], some galaxies can have log$_{10}$IRX $>$ 2.5. Our simulated galaxies located above the median relations are in a region that is known to be populated by IR bright galaxies at all redshifts \citep[e.g.][]{Casey_2014, Howell_2010, Reddy_2012, Lo_Faro_2017}.

\subsubsection{Impact of the star formation history}

The impact of the star formation history onto the IRX--$\beta$ relation has been originally studied by \cite{Kong_2004} and then discussed and confirmed by several studies \citep{Cortese_2006,Cortese_2008,Boquien_2009,Popping_2017,Narayanan_2018}: older stellar populations lead to higher intrinsic values of $\beta$. \cite{Narayanan_2018} also found that galaxies hosting older stellar populations are less obscured (the old stellar population evolves in an ISM, which is less dense than in GMC) leading to a lower IRX.

Star formation history is therefore a relevant parameter affecting the IRX--$\beta$ relation. However it is difficult to find a good tracer of the average star formation history (see \citealt{Boquien_2009} for a complete discussion about the \textit{D4000} and \textit{b} parameters). In our case we use the luminosity-weighted age of the stellar population as tracer of the star formation history.

\begin{figure}[t!]
	\begin{center}
		\includegraphics[scale=0.7]{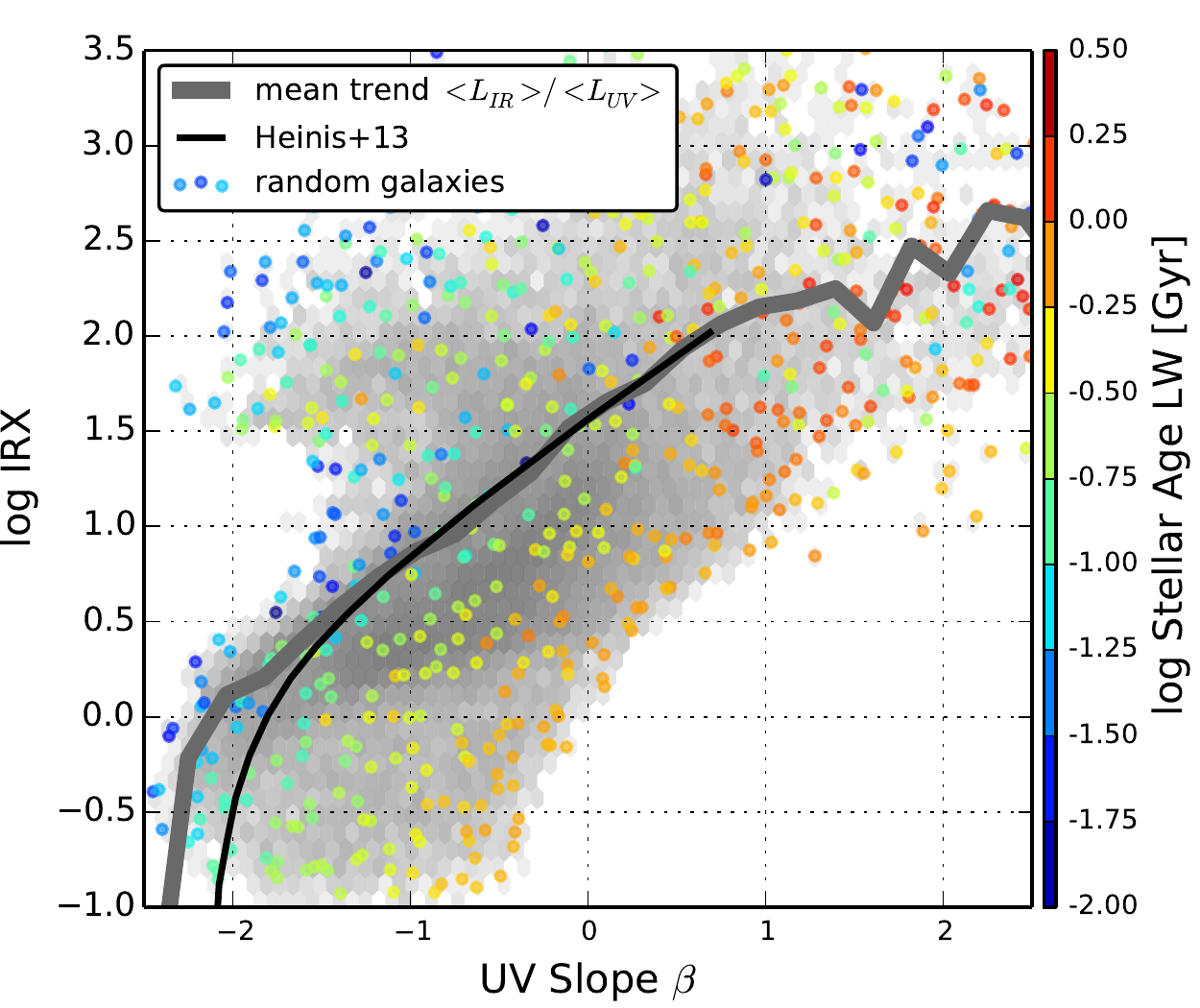}
  		\caption{IRX--$\beta$ relation for main sequence galaxies at $z = 1.5$. colored point show a random sample of star forming galaxy uniformly distributed into the IRX--$\beta$ plan. The color code is linked to the luminosity-weighted stellar population age.}
  		\label{fig:IRX_beta_age_dependency}
	\end{center}
\end{figure}

In Fig. \ref{fig:IRX_beta_age_dependency}, we highlight the luminosity-weighted age of a sub-sample of galaxies from our model uniformly selected in the IRX--$\beta$ plan. We confirm the trend found by the previous studies mentioned above: galaxies with older stellar populations preferentially lie below the canonical \cite{Meurer_1999} IRX--$\beta$ relation. 

\subsubsection{Impact of the attenuation curve}

Recent observations done with the Atacama Large Millimetre Array (ALMA) by e.g. \citealt{Capak_2015}, \citealt{Bouwens_2016}, and \citealt{Pope_2017} (z $>$ 4) or combined HST and Hershel/PACS measurements \citep{Reddy_2018} (z $>$ 1.5) have mainly populated the IRX-$\beta$ diagram below the canonical relation of \cite{Meurer_1999}. Those studies have increased the observed range of $\beta$ ($<$-1.0) at low IRX ($<$10). They concluded that such low IRX and $\beta$ values were consistent with a SMC attenuation curve (stepper than MW attenuation curve) and the low metallicity expected for the high-redshift galaxies.

\begin{figure}[t!]
	\begin{center}
		\includegraphics[scale=0.7]{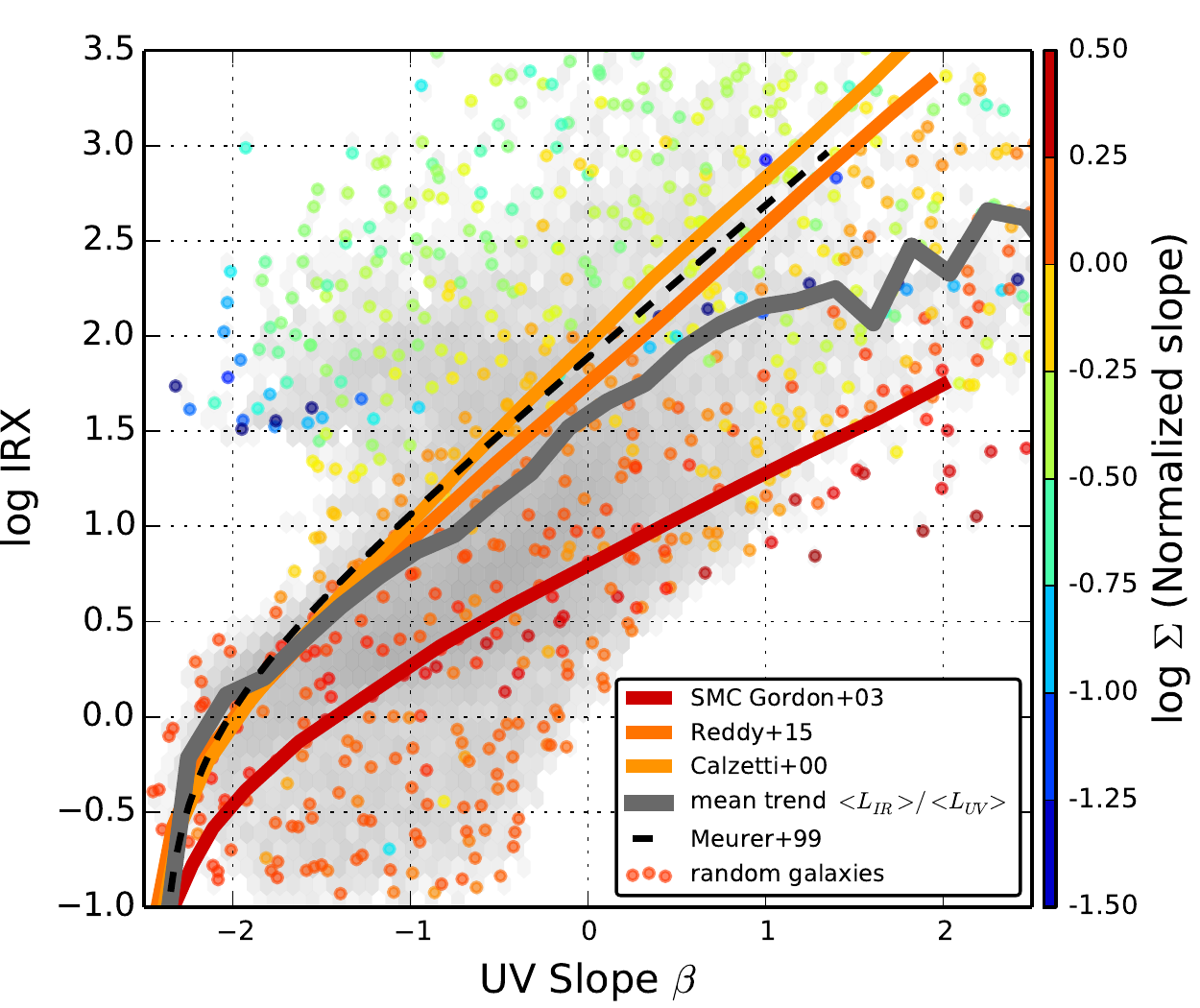}
  		\caption{IRX--$\beta$ relation for main sequence galaxies at $z = 1.5$ with colored points marking the values of the slope of the attenuation curve. The right-hand axis indicates the normalised slope of the  attenuation curve. $\Sigma$ is the ratio of the effective attenuation curve slope and the slope of the \cite{Calzetti_2000} attenuation law (here: -0.54 dex/$\mu$m). In other words, log$_{10}\Sigma >$ 0 (resp. log$_{10}\Sigma <$ 0) indicates a steeper (resp. flatter) attenuation curve than \cite{Calzetti_2000} attenuation curve. We compare with average IRX--$\beta$ computed by \cite{Salmon_2016} for attenuation curves estimated by \cite{Gordon_2003} (SMC), \cite{Calzetti_2000} and \cite{Reddy_2015}. These three average relations are plotted with solid colored lines. As for the random galaxies sample, the color code is linked to the normalized slope of the attenuation curve.}
  		\label{fig:IRX_beta_att_slope_dependency}
	\end{center}
\end{figure}

Fig.\,\ref{fig:IRX_beta_att_slope_dependency} shows the IRX--$\beta$ relation at z = 1.5 with a color coding of the attenuation curve. The slope is computed using a linear regression in the log$\lambda$--log$(A_{\lambda}/A_V)$ plan. The attenuation $A_{\lambda}/A_V$ is computed according to Eq.\,\ref{eq:effective_attenuation}. The fitting procedure is applied with seven photometric points (FUV, NUV and filters 1, 3, 5, 7 and 9 of \citealt{Calzetti_1994}). To allow for an easy comparison with the standard \cite{Calzetti_2000} attenuation law, we show in Fig.\,\ref{fig:IRX_beta_att_slope_dependency} a value normalized to the slope of the \cite{Calzetti_2000} attenuation law (-0.54 dex/$\mu$m). In other words, in Fig.\,\ref{fig:IRX_beta_att_slope_dependency} slope log$_{10}\sigma >$ 0 (resp. log$_{10}\sigma <$ 0) indicates a steeper (resp. flatter) attenuation curve than \cite{Calzetti_2000} attenuation curve.

For comparison, in addition to our stacking estimator we also include in Fig.\,\ref{fig:IRX_beta_att_slope_dependency} the average IRX--$\beta$ relations computed by \cite{Salmon_2016} and derived from attenuation curves estimated by i) \cite{Gordon_2003} (SMC), ii) \cite{Calzetti_2000} and \cite{Reddy_2015}. As shown on Fig. \ref{fig:effective_attenuation_AFUV_dependency}, slopes of these attenuation curves are different; \cite{Calzetti_2000}: -0.54dex/$\mu$m (our reference), \cite{Reddy_2015}: -0.64 dex/$\mu$m, \cite{Gordon_2003}: -1.23 dex/$\mu$m.

In Fig.\,\ref{fig:IRX_beta_att_slope_dependency}, slopes of attenuation curves associated to our simulated galaxies and trends of average IRX--$\beta$ computed by \cite{Salmon_2016} are in good agreement. We clearly observe that galaxies lying below the reference \cite{Meurer_1999} relation have steeper attenuation curves (up to three times), consistent with the SMC-like attenuation curves. Our {\tt{G.A.S.}} model is in perfect agreement with the recent studies performed by i.e. \cite{Salmon_2016}, \cite{Lo_Faro_2017}, or \cite{Reddy_2018} and with the simulations analyzed by \cite{Popping_2017} or \cite{Narayanan_2018}. 

On the contrary, galaxies lying above the reference relations have flatter attenuation curves (up to ten times, \citealt{Lo_Faro_2017}). As expected, for galaxies lying close to the reference relations, the slope of the attenuation curve is fully compatible with the one of the \cite{Calzetti_2000} sample.

As shown in Figs.\,\ref{fig:effective_attenuation_inclination_dependency} and \ref{fig:effective_attenuation_AFUV_dependency}, the {\tt{G.A.S.}} model allows us to produce a large set of effective attenuation curves agreeing with recent measurements. These effective attenuation curves can be flatter or steeper than the \cite{Calzetti_2000} law. This diversity and the variation of the slope across the IRX--$\beta$ plan is fully linked to the capability of the {\tt{G.A.S.}} model to associate with each dust contents/metallicities of each galaxy components (GMC, diffuse ISM) a dedicated extinction curve and therefore a dedicated effective attenuation.

\subsubsection{Evolution with the redshift}

\begin{figure}[t]
	\begin{center}
		\includegraphics[scale=0.66]{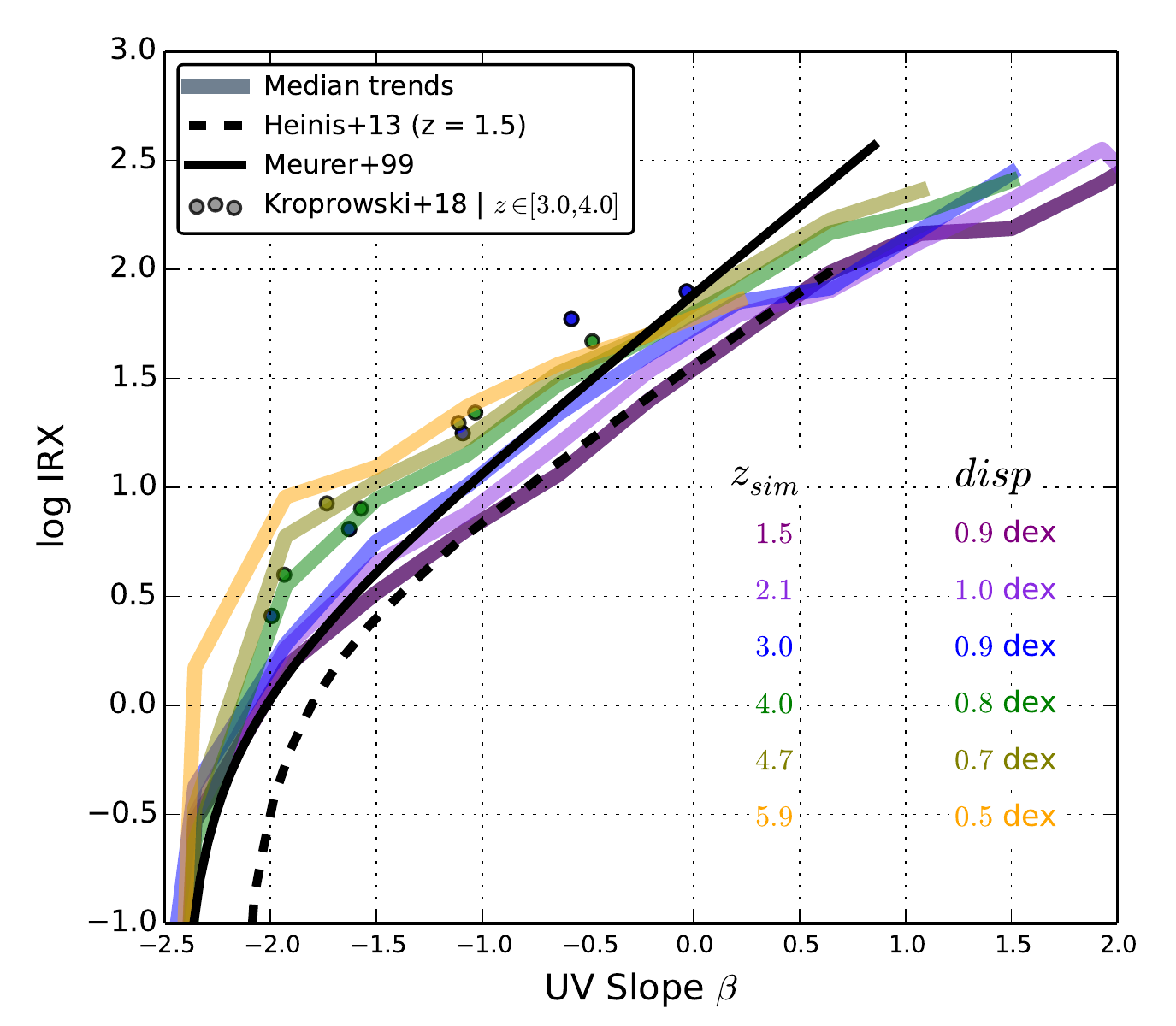}
  		\caption{Variation of the IRX--$\beta$ relation with redshift, from z = 1.5 to z $\simeq$ 6.0. For each redshift, medians trends are shown with thick coloured lines. Observational trends coming from \cite{Heinis_2013} and \cite{Meurer_1999} are shown with dashed and solid black lines respectively. Coloured points (according to the redshift colour coding) mark the recent measurements performed by \cite{Koprowski_2018} with ALMA.}
  		\label{fig:IRX_beta_redshift}
	\end{center}
\end{figure}

Fig.\,\ref{fig:IRX_beta_redshift} shows the IRX--$\beta$ relations for six different redshifts, from z = 1.5 to z $\simeq$ 6.0. We present relations built with the stacking estimator. At a given $\beta$-slope, we note an increase of the average attenuation with redshift (about 0.4\,dex between z = 1.5 to z =~6.0). Using the 15\% and 85\% percentiles, we compute the scatter of the relation that slightly decreases with the redshift, from 0.5\,dex at z $\simeq$ 6.0 to $\simeq$1.0\,dex at z = 1.5. By taking into account these scatters, the slopes and behaviors of our IRX--$\beta$ relations are compatible with the measurements, especially with \cite{Heinis_2013}. While the measurements performed with ALMA by \cite{Koprowski_2018} are above our mean relation at $z=3.35$, the measurements at $z=3.87$ and $z=4.79$ agree fairly well with our mean trend. As for the IRX--$M_{\star}$ relation, we see a progressive increase of the IRX with the redshift. This trend is difficult to validate/invalidate with current observational measurements.

%
%

\section{Conclusion}
\label{sec:discussion_conclusion}

We presented our implementation of dust extinction processes in the {\tt G.A.S.} galaxy evolution model. The effective attenuation is assumed to be caused by three types of dust grain (PAH, VSG and BG). Each galaxy in our model can host up to three components: i) an homogeneous and diffuse ISM hosting old stars, ii) some GMCs composed of fragmented/dense gas in which the youngest stars are still embedded and iii) a central bulge component hosting old stars transferred during major mergers. The proportion of the different grains is driven by the PAH mass fraction which is linked to the metallicity of each gas phase. The metallicity is computed separately in each phase according to their metal content ($m_O$)  and following the oxygen to hydrogen abundance. These different metallicities lead to different dust compositions in the different components.
In each gas phase, the evolution of the dust composition with the gas metallicity allows to compute the attenuation curves that can be shallower or steeper (SMC-like) than the extensively used \cite{Calzetti_2000} attenuation law. Even if the disk inclination affects the slope of the attenuation curve, the absolute attenuation in the FUV band appears to be the main driver. 

For each simulated galaxy, we built the intrinsic and the attenuated stellar spectrum. In each galaxy environment, the intrinsic stellar spectrum is computed following the explicit star formation history and the metal enrichment. We then produced intrinsic and extinguished UV luminosity functions, from $z\simeq 9$ to $z = 0.1$ and deduced the evolution of the absorbed fraction of UV emission. The predictions of our model are in very good agreement with measurements. We confirm that at least half of the UV radiation is absorbed and reprocessed by dust at z $\sim$ 8.0. Metal enrichment is therefore already very active at this epoch.  

By assuming that all the energy absorbed in the UV-optical wavelength range is reprocessed in the IR, we computed the IR luminosity of each galaxy. The IR luminosity function predicted by our model are in good agreement with existing observational measurements (between z = 4.0 and z = 0.3). At z = 0.1 and z = 0.4 the model slightly overestimates the total IR luminosity function, especially in the bright-end. At higher z, our model predicts that the most massive and most active star-forming galaxies can reach IR luminosities close to $10^{12.5}$, even at $z \simeq 7$. 

The analysis of the IR luminosity function reveals that the main part of the IR radiation comes from the strongly-attenuated GMCs, more than 80\% at $z>3$ and at least 50\% at $z=1.0$. The contribution coming from GMCs in the total IR emission is slightly higher in low mass/luminous IR galaxies, in which the ISM is more structured/fragmented, than in high mass/luminous IR galaxies. 

We then focused on the InfraRed Excess. For z$<$3.0, the IRX--$M_{\star}$ relation predicted by our model agrees with the recent measurements  \citep[e.g.][]{Pannella_2015,Heinis_2014,Reddy_2010}. This supports previous suggestions indicating that the stellar mass may be used as a rough proxy for dust attenuation. The choice of the $Z_g-f_{PAH}$ relation does not impact this result. At $z>3.0$, our model predicts an increase of the average IRX for a given stellar mass, although this trend is not observed (\cite{Pannella_2015,Heinis_2014,Whitaker_2014}). 

The IRX-$\beta$ correlation is one of the key tools to correct the UV emission of star forming galaxies from dust attenuation. This correlation is widely explored and discussed in the literature and large scatters and departures are found. Consistently with the observations, the dispersion of the IRX-$\beta$ relation predicted by our model is large. At a given UV slope, some galaxies hosting a strong star-formation activity ($>100 M_{\odot/yr}$) can reach IRX values 100 times larger than those predicted by the \cite{Meurer_1999} fiducial relation. 

With our model, we confirm that galaxies hosting an ``old'' ($>$ 50Myr) stellar population lie preferentially below the \cite{Meurer_1999} relation. An attenuation law steeper than the \cite{Calzetti_2000} average attenuation law also shifts galaxies below the \cite{Meurer_1999} relation. The deviations from the standard relation are essentially driven by these two factors.

This analysis confirms that $\beta$ is not a good tracer of dust attenuation, especially in galaxies hosting a strong star formation activity. The choice of the attenuation curve used to correct for attenuation is therefore crucial and has to be taken with caution.

\begin{acknowledgements}
MC thanks Laurent Verstraete for very useful discussions about the {\tt DustEM} model and Mederic Boquien for discussions about extinction prescriptions. MC thanks the Centre National d'Etude Spatial for its financial support. We acknowledge financial support from the ``Programme National de Cosmologie and Galaxies'' (PNCG) funded by CNRS/INSU-IN2P3-INP, CEA and CNES, France, from the ANR under the contract ANR-15-CE31-0017 and from the OCEVU Labex (ANR-11-LABX-0060) and the A*MIDEX project (ANR-11-IDEX-0001-02) funded by the ``Investissements d'Avenir'' French government programme managed by the ANR.
\end{acknowledgements}

\bibliographystyle{aa} 
\bibliography{GAS_II}

\newpage
\begin{appendix}

\section{PAHs}

\subsection{Impact of the $Z_g-f_{PAH}$ relation}
\label{sec:PAHs}

The $Z_g-f_{PAH}$ relation is a strong assumption for the dust attenuation model implemented in \GAS. To illustrate its impact on the effective attenuation, we created two modified relations, mostly compatible with the lowest and the highest $f_{PAH}$ values associated with the observed sample \cite{Draine_2007, Ciesla_2014, Remy_Ruyer_2015}. The two relations are defined following:
\begin{equation}
	\dfrac{f_{PAH}}{f_{PAH,0}} = 10^{-17.1 + 2.0(Z_g-0.2)}~~~~\rm{steeper ~~envelope}
	\label{eq:modified1_f_PAH}
\end{equation}
\begin{equation}
	\dfrac{f_{PAH}}{f_{PAH,0}} = 10^{-1.6 + 0.2(Z_g-0.2)}~~~~\rm{flatter~~ envelope}
	\label{eq:modified2_f_PAH}
\end{equation}
with $f_{PAH,0} = 4.57\%$, which is the reference value for our Galaxy. 

\begin{figure}[t!]
	\begin{center}
		\includegraphics[scale=0.7]{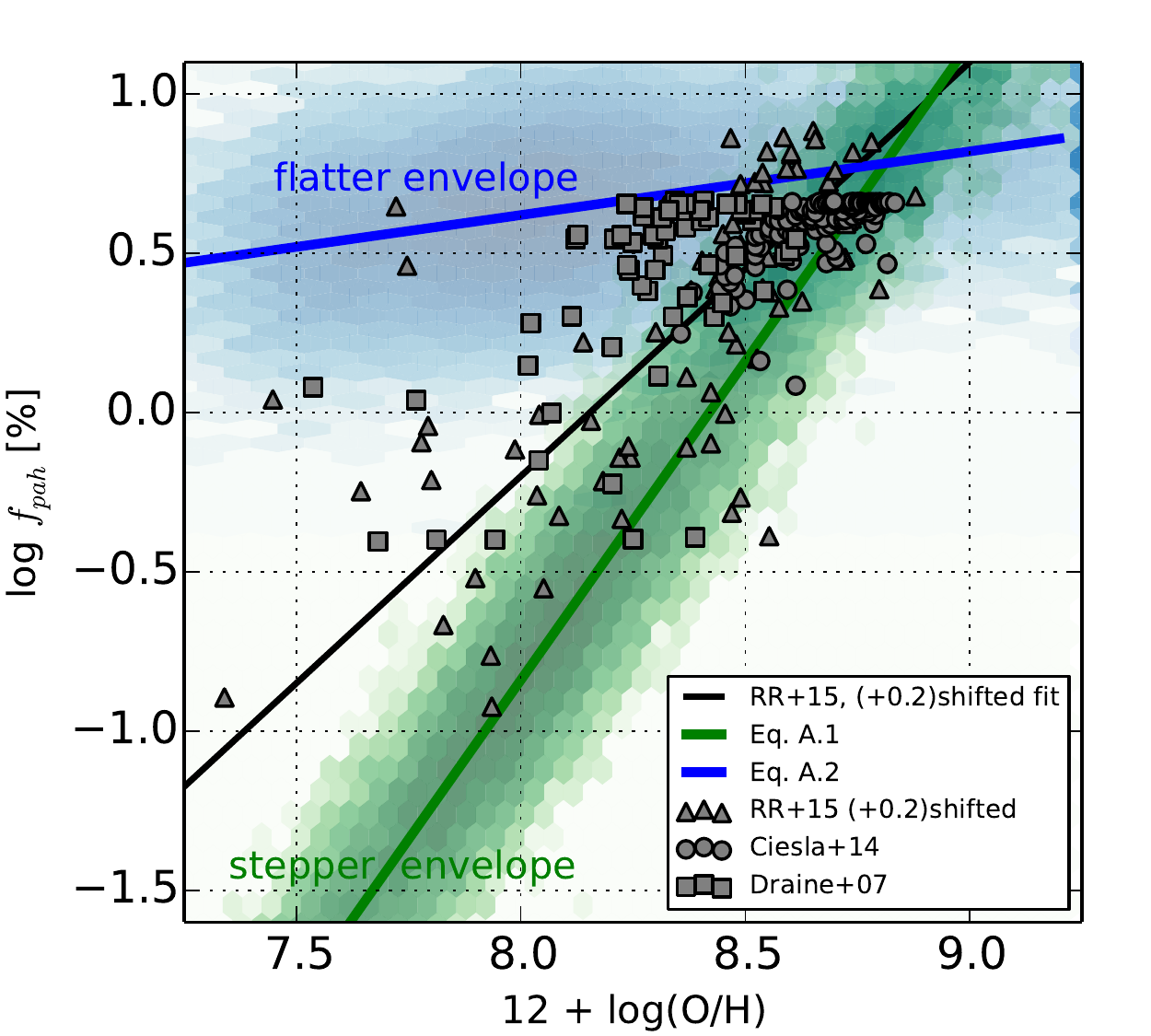}
  		\caption{Fraction of PAH as a function of the gas metallicity of the dense/fragmented gas phase. The solid blue and green lines mark the modified relations based on the flatter and the steeper relations respectively. The black solid line shows the original relation used in \GAS (black solid line in Fig.\,\ref{fig:f_pah_Zg}). The blue and the green shaded areas show the full \GAS\ distribution generated by \GAS around the two modified relations. Data points are similar to those used in Fig.\,\ref{fig:f_pah_Zg}.}
  		\label{fig:modified_f_pah_Zg}
	\end{center}
\end{figure}

In Fig.\,\ref{fig:modified_f_pah_Zg} we compare the two modified relations with the original one. According to Eq.\,\ref{eq:modified1_f_PAH} and Eq.\,\ref{eq:modified2_f_PAH} the amount of PAH is respectively reduced and increased in the gas phases. In the two cases, the dust composition of low-metallicity galaxies ($Z_g <$ 8.0) is strongly affected. 

\begin{figure}[t!]
	\begin{center}
		\includegraphics[scale=0.7]{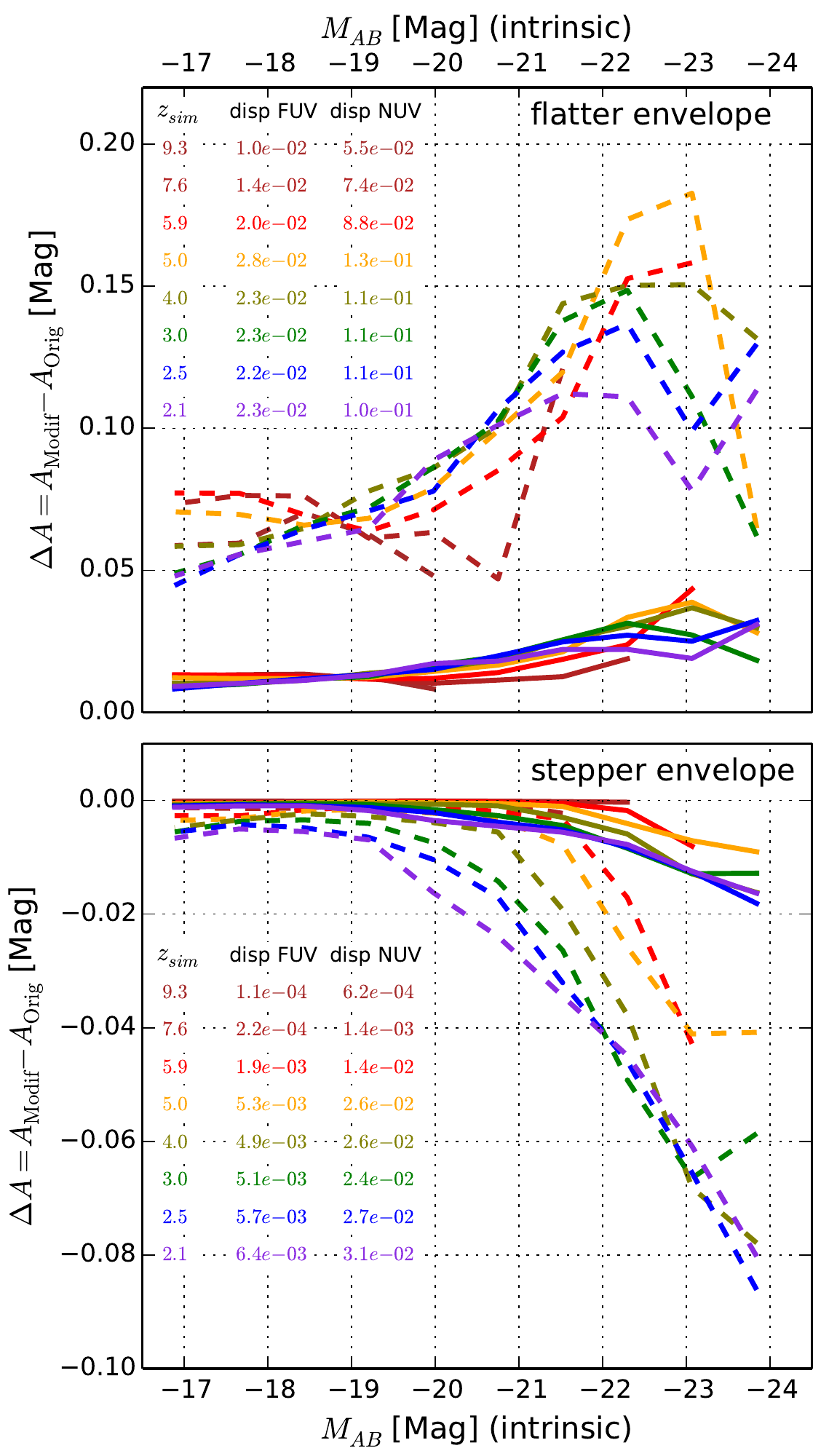}
  		\caption{Impact of the modified $Z_g-f_{PAH}$ relations onto the effective FUV and NUV attenuation (Eq. \ref{eq:effective_attenuation}). The difference between the original and the modified relation is plotted as a function of the intrinsic FUV/NUV magnitude that does not evolve between the two configurations. The upper and the lower panels show the impacts produced by the flatter limits and the stepper relations respectively. The solid and the dashed coloured lines are associated with the FUV and the NUV band respectively. $A_{\rm{Modif}}$ and $A_{\rm{Orig}}$ are the effective attenuation computed according the modified and the original $Z_g-f_{PAH}$ relations, respectively. Each coloured line marks a specific redshift from $z=9.3$ to $z=2.1$. The explicit list is given in the left side of the panels. The two other columns list the average dispersion measured for the FUV and the NUV band around the mean relation.}
  		\label{fig:dA_FUV_NUV}
	\end{center}
\end{figure}

To highlight the impact of these new $Z_g-f_{PAH}$ relations, we compare in Fig.\,\ref{fig:dA_FUV_NUV} the effective attenuation (Eq. \ref{eq:effective_attenuation}) produced by both the original ($A_{\rm{Orig}}$) and the modified ($A_{\rm{Modif}}$) relations. The analysis is performed on a sub-sample of galaxies, randomly selected at different redshifts. Attenuations are computed for both the FUV and the NUV bands. For the two bands and the two modified relations, the impact is small (only up to 0.15 Mag) and mainly affects bright galaxies. As expected, a smaller(larger) fraction of PAH leads to galaxies slightly more(less) luminous. We note that the impact is higher in the NUV than in the FUV band. This trend is fully expected. Indeed, as presented in Sects.\,\ref{sec:Slab_geometry} and \ref{sec:Dwek_geometry} the effective attenuation in the FUV band is fixed according to both the hydrogen column density and the gas phase metallicity, following \cite{Boquien_2013}. The modification of the $Z_g-f_{PAH}$ does not affect this normalisation. The fraction of PAH only slightly affects the slope of the extinction curve in the FUV band. The measured variations ($<$0.05\,Mag) are only due to this small slope change integrated through the FUV filter response. In the galaxy rest frame, the NUV band falls on the PAH bump. This bump can strongly modify the local slope of the extinction curves around the NUV band (Fig. \ref{fig:dust_ext_alb}). The impact of the new $Z_g-f_{PAH}$ relations are therefore larger in the NUV than in the FUV band. We also note that the dispersion around the mean trend is also larger in the NUV.

\begin{figure}[t!]
	\begin{center}
		\includegraphics[scale=0.7]{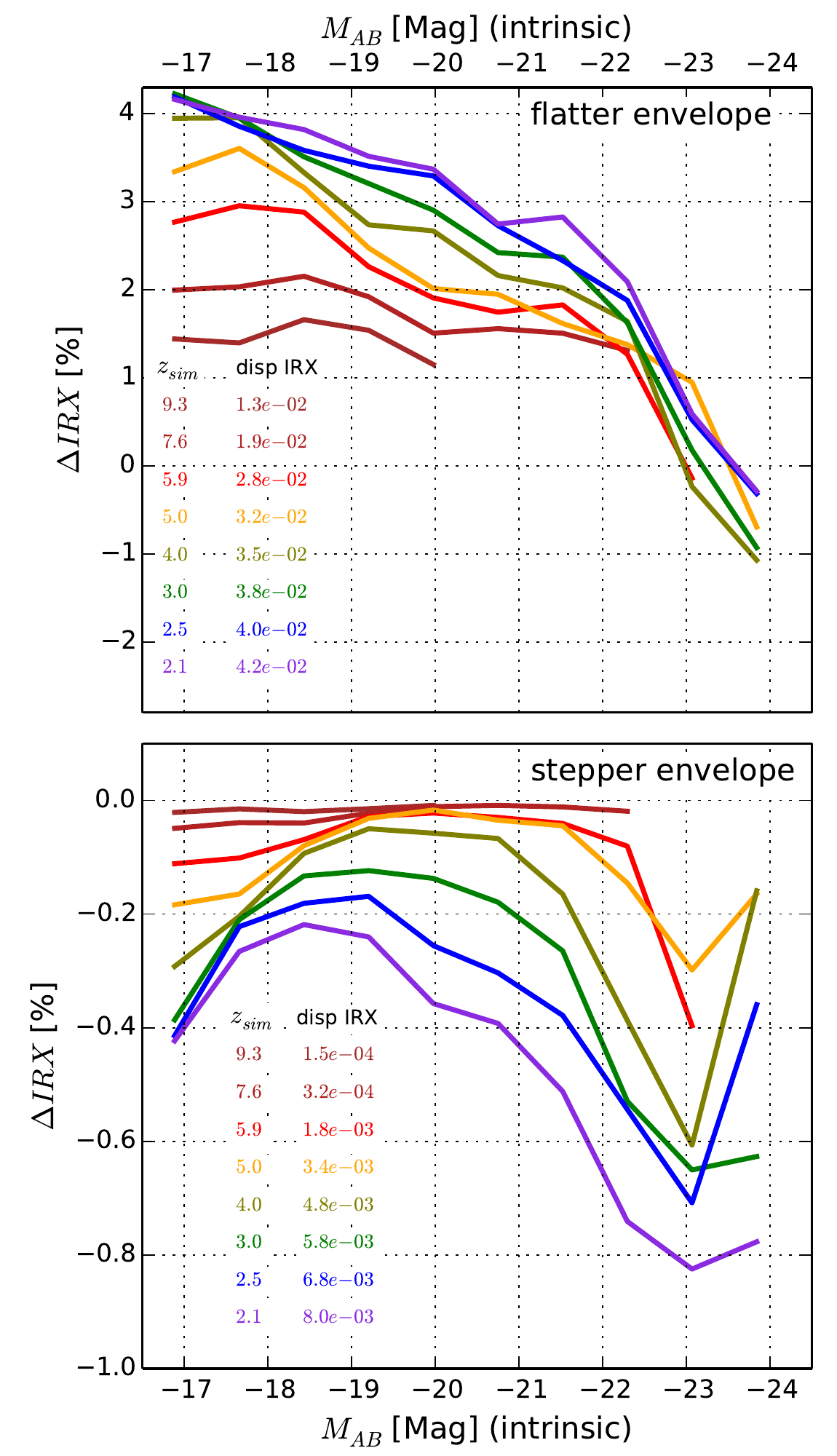}
  		\caption{Impact of the modified $Z_g-f_{PAH}$ relations on the IRX. The upper and the lower panels show the impacts produced by the upper limits and the lower limits relations flatter/steeper respectively. The difference between the original and the modified relation is plotted as a function of the intrinsic FUV magnitude that does not evolve between the two relations. Each coloured lines shows a given redhift. The list of redshift and the dispersion around the mean are indicated in the bottom left corner of each panel.}
  		\label{fig:dIRX}
	\end{center}
\end{figure}

The overall impact of these modified $Z_g-f_{PAH}$ relations can be measured through the IRX variation, as shown in Fig.\,\ref{fig:dIRX}. The variation is plotted as a function of the intrinsic FUV magnitude that does not evolve between the two relations. The impact is larger when the PAH fraction increases (up to 4\%) than when the PAH fraction decreases (only -0.8\%). By reducing(increasing) the effective amount of PAH, IRX slightly decreases(increases). As for the effective attenuation in the FUV and NUV band, higher is the redshift, smaller is the impact. In addition, UV bright galaxies are more affected than UV faint galaxies. These two trends are linked to the progressive increase of the average metallicity, from high to low redshift and from low- to high-stellar masses. 

\subsection{PAH fraction: random walk algorithm}
\label{sec:PAH_random_walk}

To allow for a smooth and continuous evolution of the PAH, VSG and BG fractions, from one time-step to the next one, the fraction of PAH evolves according to a random walk between the previous value, $f_{PAH}^{n-1}$, and a target value, $f_{PAH}^{\dag}$, expected from the empirical PAH-metallicity relation (see Eq. \ref{eq:f_PAH}): 
\begin{equation}
	f_{PAH}^{n} = f_{PAH}^{n-1} + r\times(f_{PAH}^{\dag}-f_{PAH}^{n-1}).
	\label{eq:f_PAH_n}
\end{equation}
where $r$ is a random number following a uniform distribution. For the BG component, we apply a similar algorithm. The target mass fraction is settled $f_{BG}^{\dag} = \frac{2}{3}(1 - f_{PAH}^{n})$. Then we apply:
\begin{equation}
	f_{BG}^{n} = f_{BG}^{n-1} + r\times (f_{BG}^{\dag}-f_{BG}^{n-1})
	\label{eq:f_BG_n}
\end{equation}
The mass conservation rule then requires that VSG are distributed following the residual mass.
\begin{equation}
	f_{VSG}^n = 1 - f_{PAH}^n - f_{BG}^n
	\label{eq:f_VSG}
\end{equation}

\section{The GALAKSIENN library}

The GALAKSIENN library stores the main results produced by our new \GAS\, semi-analytical model, especially MOCK galaxy catalogs and sky maps described in our paper III. It is available online through the ZENODO platform: {\tt{https://zenodo.org/}}. A complete description of the GALAKSIENN library is given in paper III. In association with this paper II, we distribute the ASCII tables of the FUV and IR luminosity functions presented in Fig. \ref{fig:UV_luminosity_functions} and Fig. \ref{fig:IR_luminosity_functions} respectively. 

\end{appendix}

\end{document}